\documentclass[12pt,preprint]{aastex}
\usepackage{epsf}

\def\linebreak{\hfil\break}

\def\
{\hfil\linebreak}

\def\deg{\ifmmode {^\circ}\else {$^\circ$}\fi}
\def\degree{\ifmmode {^\circ}\else {$^\circ$}\fi}
\def\mum{\ifmmode {\rm \mu {\rm m}}\else $\rm \mu {\rm m}$\fi}
\def\arcsec{\ifmmode ^{\prime \prime}\else $^{\prime \prime}$\fi}

\def\inch{\ifmmode ^{\prime \prime}\else $^{\prime \prime}$\fi}
\def\arcmin{\ifmmode ^{\prime}\else $^{\prime}$\fi}

\def\mearth{M$_\oplus$}
\def\msun{M$_\odot$}

\def\rsun{R$_\odot$}

\def\mstar{$M_\star$}

\def\qdstar{$Q_D^\star$}
\def\2470{[24]--[70]}

\def\q{$q$}
\def\qone{$q_1$}
\def\qtwo{$q_2$}
\def\qthree{$q_3$}
\def\qfour{$q_4$}
\def\rmax{$r_{max}$}
\def\xm{$x_m$}

\newbox\grsign \setbox\grsign=\hbox{$>$} \newdimen\grdimen \grdimen=\ht\grsign
\newbox\simlessbox \newbox\simgreatbox
\setbox\simgreatbox=\hbox{\raise.5ex\hbox{$>$}\llap
     {\lower.5ex\hbox{$\sim$}}}\ht1=\grdimen\dp1=0pt
\setbox\simlessbox=\hbox{\raise.5ex\hbox{$<$}\llap
     {\lower.5ex\hbox{$\sim$}}}\ht2=\grdimen\dp2=0pt

\def\simless{\mathrel{\copy\simlessbox}}

\begin{document}

\title{Coagulation Calculations of Icy Planet Formation at 15--150 AU: A Correlation Between 
the Maximum Radius and the Slope of the Size Distribution for Transneptunian Objects}
\vskip 7ex
\author{Scott J. Kenyon}
\affil{Smithsonian Astrophysical Observatory,
60 Garden Street, Cambridge, MA 02138} 
\email{e-mail: skenyon@cfa.harvard.edu}

\author{Benjamin C. Bromley}
\affil{Department of Physics, University of Utah, 
201 JFB, Salt Lake City, UT 84112} 
\email{e-mail: bromley@physics.utah.edu}
%
%

\begin{abstract}

We investigate whether coagulation models of planet formation can explain the 
observed size distributions of transneptunian objects (TNOs).  Analyzing 
published and new calculations, we demonstrate robust relations between the 
size of the largest object and the slope of the size distribution for sizes 
0.1~km and larger. These relations yield clear, testable predictions for TNOs 
and other icy objects throughout the solar system.  Applying our results to 
existing observations, we show that a broad range of initial disk masses,
planetesimal sizes, and fragmentation parameters can explain the data. Adding 
dynamical constraints on the initial semimajor axis of `hot' KBOs along with
probable TNO formation times of 10--700~Myr restricts the viable models to 
those with a massive disk composed of relatively small (1--10~km) planetesimals.

\end{abstract}


\section{INTRODUCTION}
\label{sec: intro}

The transneptunian objects (TNOs, small icy planets orbiting the Sun beyond Neptune)
provide crucial tests of planet formation theories. With its rich dynamical structure, 
the current orbital architecture of TNOs constrains models for the dynamical interactions
among newly-formed gas giants and for the origin of the Oort cloud \citep[e.g.,][]{MorbiSSBN}.
The diverse populations of binary TNOs provide additional constraints on the dynamical
history of the solar system \citep{noll2008,nesv11,murrayclay2011}. Finally, the sizes of the 
largest TNOs and the size distributions within the dynamical families of TNOs constrain 
the formation histories of the gas giants, the TNOs, and perhaps the terrestrial planets 
\citep[e.g.,][]{kl1998,kenyon2002,gomes2005,MorbiSSBN}.

There are two broad concepts for the formation and evolution of TNOs. 
Both begin with small dust grains within a massive gaseous disk surrounding a young star. 
As the disk evolves, dust grains collide, merge, and grow into mm- to cm-sized objects 
which decouple from the gas and fall into the midplane of the disk 
\citep[e.g.,][and references therein]{birn2010}.
Once a large fraction of the solid material is in the midplane, the two paths to TNOs diverge.
In one path, icy objects continue to collide, merge, and grow into roughly km-sized 
planetesimals and then into much larger TNOs \citep[e.g.,][]{weiden1997a}.
In the other path, instabilities or turbulent eddies in the disk concentrate icy objects into 
massive clumps which collapse directly into large TNOs 
\citep[e.g.,][]{johansen2007,chambers2010,cuzzi2010,pan2011}. Once TNOs 
form, the evolutionary paths converge: dynamical interactions with the gas giants produce 
the diverse architecture of the current TNO population \citep[e.g.,][]{MorbiSSBN}.

Testing the two formation paths requires robust theoretical models which make clear 
predictions for the maximum radius and the size distribution of TNOs.  At present, 
analytic theory and numerical simulations cannot predict the outcome of instability 
mechanisms \citep{chiang2010,pan2011}. However, coagulation models yield robust predictions 
for the time evolution of the size distribution and the sizes of the largest objects
\citep[e.g.,][and references therein]{kl1999a,kenyon2002,kb2004c,kbod2008}. Thus, it 
is possible to compare the results of coagulation calculations with the observable 
properties of TNOs \citep{kl1999b,kb2004c,fraser2009b}.

To test the coagulation model in detail, we analyze a set of published \citep{kb2008, kb2010} 
and new coagulation calculations. Our numerical models follow the collisional and dynamical 
evolution of small planetesimals into TNOs at orbital semimajor axes $a$ = 15--150~AU for 
evolution times 1--10~Gyr. This range of semimajor axes lies outside the most likely region
of giant planet formation \citep[e.g.,][]{raf2004,gold2004,kk2008,kb2009}.
By considering a broad range of initial planetesimal sizes 
(and size distributions) and a range of fragmentation parameters, we derive the time evolution 
of the radius of the largest object and the slope of the size distribution as a function of 
initial conditions and distance from the Sun.  Comparisons of our results with the observations 
allows an assessment of the match between the observed properties of TNOs and the models.

Coagulation models yield robust predictions for the relation between the size of the largest 
object and the slope of the size distribution. Although this relation is insensitive to 
distance from the Sun and fairly insensitive to the fragmentation parameters, the evolution 
depends on the initial disk mass, the initial size of the largest planetesimals, and the 
initial size distribution of planetesimals. 

Using recent observations from large surveys of TNOs, we demonstrate that collisional evolution 
often yields size distributions similar to those observed. On their own, the observations favor
models with 1--10~km planetesimals relative to those with 100~km planetesimals. Adding dynamical 
constraints on the origin of the hot and cold populations of KBOs to our analysis, we show that 
models in (i) massive disks with 1~km planetesimals and (ii) low mass disks with 10~km 
planetesimals provide equally satisfactory matches to the data. Requiring that TNOs form prior 
to the Late Heavy Bombardment eliminates models with low mass disks. Thus, all of the available 
data taken together imply that TNOs formed in a massive disk composed of 1~km planetesimals.

We begin our discussion with an overview of the numerical code in \S\ref{sec: calcs}.
After describing basic results of the calculations in \S\ref{sec: results}, we compare
the results with observations in \S\ref{sec: apps}. We conclude with a brief summary in 
\S\ref{sec: summary}.

\section{PLANET FORMATION CALCULATIONS}
\label{sec: calcs}

To calculate the formation and evolution of TNOs, we use a hybrid 
multiannulus coagulation--$N$-body code.  We compute the collisional evolution of an 
ensemble of planetesimals in a circumstellar disk orbiting a central star of mass \mstar. 
The code uses statistical algorithms to evolve the mass and velocity distributions 
of low mass objects with time and an $N$-body algorithm to follow the individual 
trajectories of massive objects. \citet{kb2008} describe the statistical (coagulation) 
code; \citet{bk2006} describe the $N$-body  code. Here, we briefly summarize the basic 
aspects of our approach.

We perform calculations on a cylindrical grid with inner radius $a_{in}$ and outer
radius $a_{out}$ surrounding a star with \mstar\ = 1 \msun. The model grid contains 
$N$ concentric annuli with widths 
$\delta a_i = 0.025 a_i$ centered at semimajor axes $a_i$. Calculations begin with a 
cumulative mass distribution $n_c(m_{ik}) \propto m_{ik}^{-q_{init}^{\prime}}$ of 
planetesimals with mass density $\rho_p$ = 1.5 g cm$^{-3}$ and maximum initial mass 
$m_0$.  Although our code explicitly tracks the mass distribution of solid objects, 
in the rest of the text we describe initial conditions and results in terms of 
cumulative size distributions, $n_c(r)$. For a power law cumulative size distribution,
$n_c \propto r^{-q}$, the power-law exponent is a factor of 3 larger than the exponent
for the mass distribution. Thus, $q_{init} = 3 q_{init}^{\prime}$. For this paper,
we adopt $q_{init}$ = 0.5 (most of the mass in the largest planetesimals) and 3.0
(mass distributed equally per logarithmic interval in mass/radius).

Planetesimals have horizontal and vertical velocities $h_{ik}(t)$ and $v_{ik}(t)$ 
relative to a circular orbit.  The horizontal velocity  is related to the orbital 
eccentricity, $e_{ik}^2(t)$ = 1.6 $(h_{ik}(t)/V_{K,i})^2$, where $V_{K,i}$ is the 
circular orbital velocity in annulus $i$.  The orbital inclination depends on the 
vertical velocity, $i_{ik}^2(t)$ = sin$^{-2}(2(v_{ik}(t)/V_{K,i}))$.

The mass and velocity distributions of the planetesimals evolve in time due to
inelastic collisions, drag forces, and gravitational encounters.  As summarized 
in \citet{kb2004a,kb2008}, we solve a coupled set of coagulation equations which
treats the outcomes of mutual collisions between all particles with mass $m_j$ 
in annuli $a_i$.  We adopt the particle-in-a-box algorithm, where the physical 
collision rate is $n \sigma v f_g$, $n$ is the number density of objects, 
$\sigma$ is the geometric cross-section, $v$ is the relative velocity, and 
$f_g$ is the gravitational focusing factor 
\citep{saf1969,liss1987,spaute1991,weth1993,weiden1997b,kl1998,kb2008}.  For a specific 
mass bin, our solutions include terms for (i) loss of mass from gas drag and mergers
with other objects and (ii) gain of mass from collisional debris and mergers of 
smaller objects.

Aside from deriving the cross-section and relative collision velocity, the main 
ingredient in this calculation is the gravitational focusing factor for a collision
between two planetesimals with masses (radii) $m_i$ and $m_j$ ($r_i$ and $r_j$). 
In the high velocity, dispersion-dominated regime, we adopt a variant of the piecewise 
analytic approximation of \citet[][see also Kenyon \& Luu 1998]{spaute1991}:
\begin{equation}
f_{g,d} = \left\{
\begin{array}{lll}
1 + \beta (v_e/v)^2  & \hspace{5mm} & v > 0.32 v_e , \\
42.4042 ~ (v_e/v)^{1.2} & & 0.01 < v / v_e \le 0.32 , \\
10651.453 & & v / v_e \le 0.01 , \\
\end{array}
\right.
\label{eq:fg-disp}
\end{equation}
where $\beta$ is an order unity constant which accounts for the distribution of impact 
velocities \citep{green1992}, $v$ is the velocity of a planetesimal, and $v_e$ is the 
escape velocity of the merged pair of planetesimals.

For the low velocity, shear-dominated regime, we define 
the angular velocity $\Omega = (G M_\star / a^3)^{1/2}$,
the Hill radius $r_H = a (m / 3 M_\odot)^{1/3}$,
the Hill velocity $v_H = \Omega r_H$, and
an angular scale factor $\psi = (\rho_\odot / \rho_p)^{1/3}$ \rsun/$a$, where 
$\rho_\odot$ is the mean density of the Sun and \rsun\ is the solar radius.  
In the \citet{green1991} approach, 
\begin{equation}
f_{g,s} = \left\{
\begin{array}{lll}
(1 + v_e^2/v_T^2) (v_T / v) (r_H / r_T) & \hspace{5mm} & v \lesssim v_H , \\
0.5 (1 + v_e^2/v_T^2)^{1/2} (v_T / v) (r_H / r_T) (H / (r_i + r_j)) & & v \lesssim v_H ~,~ v \lesssim \psi^{1/2} v_H \\
\end{array}
\right.
\label{eq:fg-shear-t}
\end{equation}
where $r_T = a ((m_i + m_j) / m_\star )^{2/5}$ is the Tisserand radius,
$v_T = 1.1 \Omega \delta a_T$ is the Tisserand velocity, and $H$ is the 
vertical scale height of the planetesimals. In the \citet{gold2004} approach,
\begin{equation}
f_{g,s} = \left\{
\begin{array}{lll}
\psi^{-1} (v_h / v) & \hspace{5mm} & v \lesssim v_H , \\
\alpha^{-3/2} & & v \lesssim v_H ~,~ v \lesssim \psi^{1/2} v_H \\
\end{array}
\right.
\label{eq:fg-shear-h}
\end{equation}
Tests using both approaches yield similar results for the collision rate.  The
simpler \citet{gold2004} relations are computationally cheaper. For either approach, 
we employ a simple algorithm to affect a smooth transition in $f_g$ between the 
dispersion and shear rates \citep{kl1998}. Because $\psi$ decreases with $a$, icy
planet formation at 15--150 AU is less likely to enter the shear-dominated regime
than rocky planet formation at 1 AU.

Collision outcomes depend on the ratio $Q_c/Q_D^*$, where $Q_D^*$ is the collision 
energy needed to eject half the mass of a pair of colliding planetesimals to infinity
and $Q_c$ is the center of mass collision energy. For two colliding planetesimals with 
masses $m_1$ and $m_2$, the mass of the merged planetesimal is
\begin{equation}
m = m_1 + m_2 - m_{ej} ~ ,
\label{eq:msum}
\end{equation}
where the mass of debris ejected in a collision is
\begin{equation}
m_{ej} = 0.5 ~ (m_1 + m_2) \left ( \frac{Q_c}{Q_D^*} \right)^{9/8} ~ .
\label{eq:mej}
\end{equation}
To place the debris in our grid of mass bins, we set the mass of the largest collision
fragment as $m_L = 0.2 m_{esc}$ and adopt a cumulative size distribution
$n_c \propto r^{-q_d}$ \citep{davis1985}. Here, we adopt $q_d$ = 2.5, the 
value expected for a system in collisional equilibrium 
\citep{dohn1969,will1994,tanaka1996,obrien2003,koba2010}. 

This approach allows us to derive ejected masses for catastrophic collisions
with $Q_c \sim Q_D^*$ and for cratering collisions with $Q_c \ll Q_D^*$
\citep[see also][]{weth1993,will1994,tanaka1996,stcol1997a,kl1999a,obrien2003,koba2010}. 
Consistent with N-body simulations of collision outcomes 
\citep[e.g.,][]{benz1999,lein2008,lein2009}, 
we set
\begin{equation}
Q_D^* = Q_b r^{\beta_b} + Q_g \rho_p r^{\beta_g}
\label{eq:Qd}
\end{equation}
where $Q_b r^{\beta_b}$ is the bulk component of the binding energy,
$Q_g \rho_g r^{\beta_g}$ is the gravity component of the binding energy,
and $r$ is the radius of a planetesimal.

\begin{figure}
\includegraphics[width=6.5in]{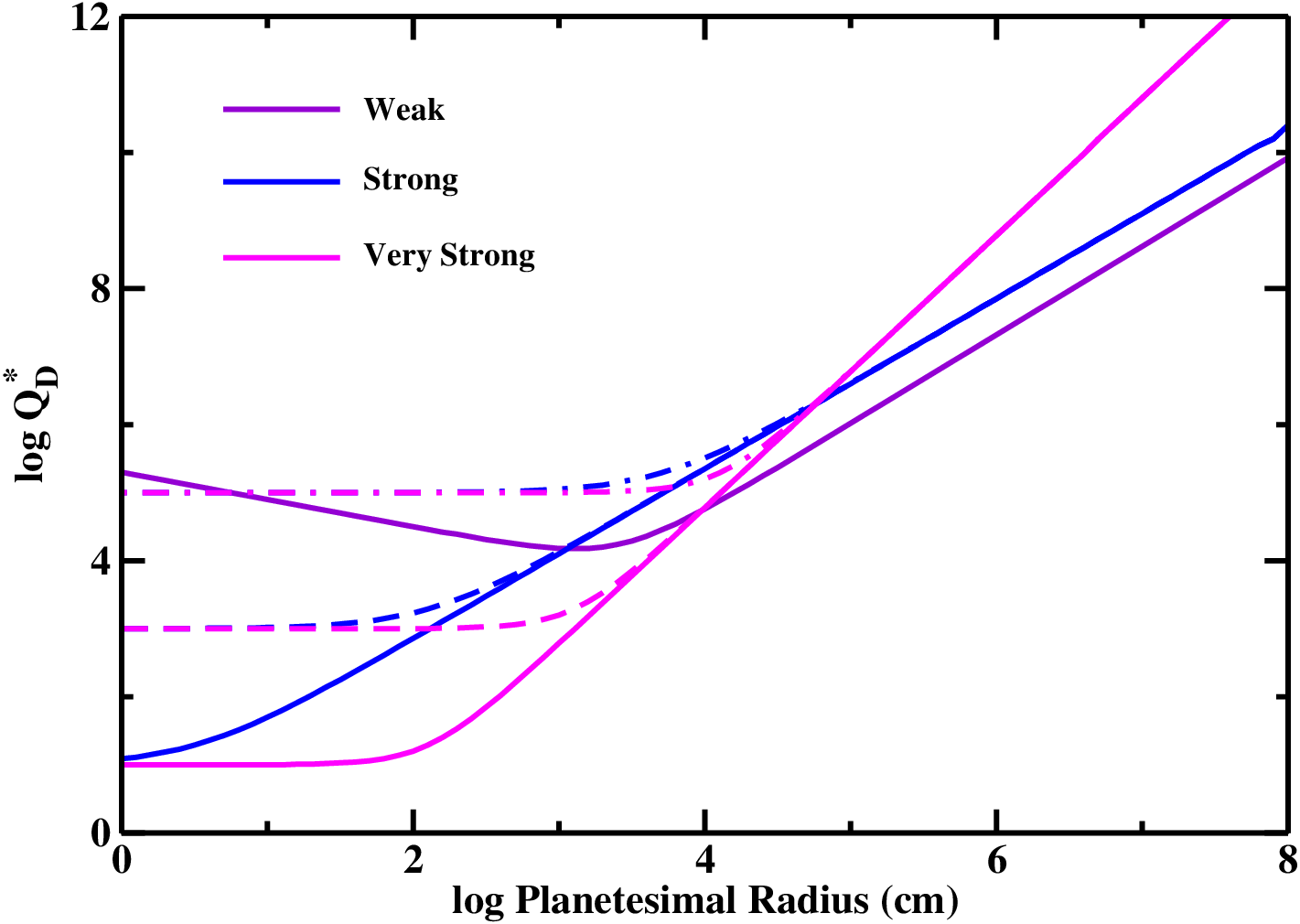}
\vskip 3ex
\caption{%
Variation of \qdstar\ with radius for the three sets of fragmentation parameters adopted in 
the text.  For the strong and very strong planetesimals, the solid ($Q_b$ = 10), dashed 
($Q_b = 10^3$) and dot-dashed curves ($Q_b = 10^5$) indicate a range of parameters for the 
strength regime.
\label{fig: qstar}
}
\end{figure}

To explore the sensitivity of our results to the fragmentation algorithm, we consider 
three sets of parameters $f_i$ (Fig. \ref{fig: qstar}). As in \citet[][2010]{kb2008}, 
strong planetesimals have $f_s$ = \citep[$Q_b$ = $10^1$, $10^3$, or 
$10^5$ erg g$^{-1}$, $\beta_b$ = 0, $Q_g$ = 2.25 erg g$^{-2}$ cm$^{1.75}$, 
$\beta_g$ = 1.25;][]{benz1999}; weaker planetesimals have $f_w$ =
\citep[$Q_b$ = $2 \times 10^5$ erg g$^{-1}$ cm$^{0.4}$, $\beta_b = -0.4$,
$Q_g$ = 0.33 erg g$^{-2}$ cm$^{1.7}$, $\beta_g$ = 1.3;][]{lein2009}.
Here, we also consider a set of very strong planetesimals with
$f_{vs}$ = \citep[$Q_b$ = $10^1$, $10^3$, or $10^5$ erg g$^{-1}$, $\beta_b$ = 0,
$Q_g$ = $10^{-4}$ erg g$^{-2}$ cm, $\beta_g$ = 2.0; e.g.,][]{davis1985}.
At small sizes, these parameters are broadly consistent with results from 
laboratory experiments of impacts with icy targets and projectiles 
\citep[e.g.,][]{ryan1999,arakawa2002,giblin2004,burchell2005}.

For our calculations, these parameters have several consequences. When fragmentation
begins, most of the mass is in planetesimals with $r \approx$ 1--100~km.  With 
\qdstar($r$ = 1--100~km) $>$ \qdstar($r$ = 1--$10^5$~cm), the gravity regime of the
fragmentation law sets the collision velocity for the onset of fragmentation.  To 
achieve catastrophic fragmentation with $Q_c/Q_D^*$ = 1, weak planetesimals require
the smallest collision velocity; very strong planetesimals require the largest. In 
all cases, once large objects begin to fragment, smaller objects also fragment. Thus,
the onset of fragmentation for 1--100 km planetesimals initiates a collisional cascade 
where all small objects are ground to dust.

To compute the evolution of the velocity distribution, we include collisional
damping from inelastic collisions, gas drag, and gravitational interactions.
For inelastic and elastic collisions, we follow the statistical, Fokker-Planck
approaches of \citet{oht1992} and \citet{oht2002}, which treat pairwise interactions
(e.g., dynamical friction and viscous stirring) between all objects in all annuli.
As in \citet{kb2001}, we add terms to treat the probability that objects in annulus
$i_1$ interact with objects in annulus $i_2$ \citep[see also][KB08]{kb2004b}. We also
compute long-range stirring from distant oligarchs \citep{weiden1989}. For gas drag,
we derive velocity damping and the inward drift of material in the quadratic, 
Stokes, and Epstein limits \citep[e.g.,][]{ada76, weiden1977a, raf2004}.

To evolve the gas in time, we consider a simple nebular model for the gas density.
We adopt a scale height $H_{gas}(a) = H_{gas,0} (a/a_0)^{1.125}$ \citep{kh1987}
and assume the gas surface density declines exponentially with time
\begin{equation}
\Sigma_{gas}(a,t) = \Sigma_{gas,0} ~ x_m ~ a^{-n} ~ e^{-t/t_{gas}}
\label{eq:sigma-gas}
\end{equation}
where $\Sigma_{gas,0}$ and $x_m$ are scaling factors and $t_{gas}$ = 10~Myr is the 
gas depletion time. Although longer than the 2--5~Myr timescale estimated from 
observations of the lifetimes of accretion disks in pre-main sequence stars 
\citep{currie2009,kk2009,mama2009,wil2011}, shorter gas depletion times have little impact 
on our results.

In the $N$-body code, we directly integrate the orbits of objects with masses larger
than a pre-set `promotion mass' $m_{pro}$. The calculations allow for mergers among
the $N$-bodies. Additional algorithms treat mass accretion from the coagulation grid 
and mutual gravitational stirring of $n$-bodies and mass batches in the coagulation 
grid. For the applications in this paper, the few large objects capable of promotion 
into the $N$-body code never contain a significant fraction of the mass in an annulus 
and never contribute significantly to the local stirring. To treat situations where a 
few large objects might impact the evolution, we set $m_{pro} = 10^{26}$ g. However, 
our calculations never produced more than a few $N$-bodies which remain on nearly
circular orbits throughout their evolution.

The initial conditions for these calculations are appropriate for a disk with an age 
of $\lesssim$ 1--2~Myr \citep[e.g.,][and references therein]{wil2011}.  We consider 
systems of $N = 64$ annuli in disks where the initial surface density of solid material 
follows a power law in semimajor axis,
\begin{equation}
\Sigma_{d,i} = \Sigma_{d,0}(M_{\star}) ~ x_m ~ a_i^{-n} ~ , 
\label{eq:sigma-dust}
\end{equation}
where $a_i$ is the central radius of the annulus in~AU, $n$ = 1 or 3/2, and
$x_m$ is a scaling factor.  For a standard gas to dust ratio of 100:1, 
$\Sigma_{gas,0} = 100 ~ \Sigma_{d,0} (M_{\star})$.  To explore a range of disk masses 
similar to the observed range among the youngest stars, we consider 
$\Sigma_{d,0}$ = 30~g~cm$^{-2}$ and $x_m$ = 0.01--3. 
Disks with $x_m \approx$ 0.1 have masses similar to the median disk masses observed around 
young stars in nearby dark clouds \citep{and2005,wil2011}. Somewhat larger scale factors,
$x_m \approx$ 1, correspond to models of the minimum mass solar nebula models of 
\citet{weiden1977b} and \citet{hayashi1981}. 

Table \ref{tab: modgrid} summarizes the model grids. For each set of $x_m$, we choose the 
extent of the disk, $a$ = 15--75 AU or 30--150 AU; initial radius of the largest planetesimal, 
$r_0$ = 1, 10, 100~km; the initial eccentricity $e_0 = 10^{-4}$ and inclination
$i_0 = 5 \times 10^{-5}$; the power law exponent of the initial size distribution, 
$q_{init}$ = $3$ (equal mass per log interval in mass) or $q_{init} = 0.5$ (most of 
the mass in the largest objects);
the fragmentation parameters, $f_s$, $f_{vs}$, or $f_w$; and the evolution time $t_{max}$.
To understand the possible range of outcomes, we repeat calculations 5--15 times for each 
combination of initial conditions. \citet[][2010]{kb2008} discuss some aspects of these 
calculations in the context of observations of debris disks.

Our calculations follow the time evolution of the mass and velocity distributions of 
objects with a range of radii, $r_{ij} = r_{min}$ to $r_{ij} = r_{max}$.  The upper 
limit $r_{max}$ is always larger than the largest object in each annulus.  To save 
computer time, we do not consider small objects which do not significantly affect the 
dynamics and growth of larger objects, $r_{min}$ = 100 cm.  Erosive collisions produce 
objects with $r_{ij}$ $< r_{min}$ which are `lost' to the model grid. Lost objects are 
more likely to be ground down into smaller objects than to collide with larger objects
in the grid \citep[see][2002b, 2004a]{kb2002a}.

\section{EVOLUTION OF THE SIZE DISTRIBUTION}
\label{sec: results}

At the start of each calculation, dynamical processes drive the evolution.  Gas drag damps 
the velocities of the smallest objects.  Dynamical friction damps the orbital velocities 
of the largest objects; dynamical friction and viscous stirring excite velocities of the 
smallest objects \citep[][2010]{gold2004,kb2008}. With stirring timescales much faster than
growth timescales, it takes 10--50 orbits to achieve a relaxed distribution of velocities
among all planetesimal sizes.  For calculations with $r_0$ = 1--10~km, this evolution occurs 
in the dispersion-dominated regime; dynamics and growth are slow and orderly. When $r_0$ = 
100~km, dynamical evolution rapidly drives the system from the shear regime to the dispersion 
regime. Thus, initial growth rates for large planetesimals are also slow and orderly. 

As the evolution proceeds, dynamical processes produce substantial gravitational focusing 
factors. Runaway growth begins. The largest objects then grow much more rapidly than, and 
`run away' from, somewhat smaller protoplanets. The initial stages of this process are in 
the dispersion regime; accretion in the shear regime dominates the later stages. Throughout
this evolution, the largest objects -- with radii of 300--1000~km -- continue to stir the 
smallest objects.  Stirring reduces gravitational focusing factors, ending shear-dominated
accretion and runaway growth.  Oligarchic growth in the dispersion regime begins.

Throughout oligarchic growth, protoplanets evolve in two distinct ways. The largest
protoplanets -- known as oligarchs -- accrete material primarily from the reservoir 
of small planetesimals. Because gravitational focusing factors are small, growth is
orderly.  Small oligarchs grow faster than large oligarchs.  Both grow faster than 
smaller planetesimals \citep[e.g.,][]{gold2004}. As all of the oligarchs grow, they 
stir the smallest planetesimals to larger and larger orbital velocities.  Eventually, 
collisions among small planetesimals become destructive, producing smaller 
planetesimals and some debris. Collisions among smaller planetesimals are even more 
destructive, leading to a collisional cascade where small planetesimals are ground to 
dust \citep[e.g.,][and references therein]{dohn1969,will1994,koba2010,bely2011}.
By robbing the oligarchs of small planetesimals, the collisional cascade limits the
growth of the largest planets to $\sim 3 -10 \times 10^3$~km 
\citep[e.g.,][]{inabawet2003,kb2008,kb2010,koba2010}.

On 1--10~Gyr time scales, the collisional cascade removes most of the initial mass in solids 
at $a \lesssim$ 100 AU \citep{kb2008, kb2010}. \citet{kb2008} express the fraction of solid 
material remaining in the disk as $f_s \approx f_0 x_m^{-1/4} (a / {\rm 30~AU})^{5/4}$. 
Collisional growth proceeds more rapidly at smaller $a$ in more massive disks. For a fixed
evolution time, the cascade removes more mass at smaller $a$ and less mass in less massive
disks.  For calculations with $r_0$ = 1~km and strong (weak) planetesimals, $f_{0,s} \approx$ 
0.12 (0.08) at 1~Gyr; $f_{0,s} \approx$ 0.10 (0.07) at 10~Gyr \citep{kb2008,kb2010}.  For the 
very strong planetesimals considered here, $f_{0,vs} \approx$ 0.26 (0.16) at 1~Gyr (10~Gyr).  
Thus, the cascade removes 74\% to 93\% of the initial mass at 30~AU on 1--10~Gyr timescales. 

The amount of solid material remaining in the disk depends on the initial maximum size of the
planetesimals. In calculations with $r_0$ = 10~km, the cascade removes roughly a factor of two 
less mass over 1--10 Gyr. When $r_0$ = 100~km, the cascade typically removes a factor of 7--8 
less mass.  Effective cascades require small planetesimals \citep{kb2010}. Ineffective cascades 
leave more mass in the disk and allow more massive planets to form. Thus, the most massive planets 
grow in ensembles of large and very strong planetesimals \citep[see also][]{kb2008,kb2010}.

Fig. \ref{fig: sd1} illustrates the outcomes of growth for an ensemble of 1~km planetesimals
in eight annuli 
at 27--32~AU. Our starting conditions place most of the mass in 1~km planetesimals; the 
cumulative size distribution of planetesimals is nearly flat. During orderly growth,
$\sim 10^4$ planetesimals reach radii of nearly 10~km in 3~Myr. Once runaway growth begins,
a handful of planetesimals grow as large as 100~km. In both cases, the size distribution
follows a shallow power law at small radii and an exponential at large radii 
\citep[e.g.,][]{davis1999b,kl1999b}.  
As the evolution makes the transition from runaway to oligarchic growth, the largest
objects grow from $\sim$ 100~km (10 Myr) to $\sim$ 500~km (30 Myr) to $\sim$ 2000~km
(300~Myr).  Because the oligarchs grow faster than smaller planetesimals, the size 
distribution becomes a fairly smooth power law from $\sim$ 10 km to $\sim$ 2000~km
\citep[e.g.,][]{kb2004c,fraser2009b,schlichting2011}.

At small sizes ($r \lesssim$ 10~km), the collisional cascade produces inflection points 
in the size distribution \citep{kb2004c,pan2005,benavidez2009,fraser2009b}. Initially, 
the cascade `erodes' 
planetesimals with $r \lesssim$ 1~km.  Smaller planetesimals erode more than large 
planetesimals. These planetesimals grow smaller with time. Because the cascade removes
smaller planetesimals from an annulus more rapidly than larger planetesimals, the size
distribution becomes shallower, creating an inflection point where erosion dominates
growth. Eventually, larger collision velocities cause small planetesimals to shatter  
completely on impact. This process creates another inflection point where shattering
dominates erosion. At the smallest sizes, debris from collisions of larger planetesimals 
dominates shattering, resulting in a sharp rise in the size distribution 
\citep[e.g.,][]{kb2004c,kb2005,pan2005,fraser2009b}.

To analyze the time-evolution of the size distribution for a variety of initial conditions,
we perform a series of least-squares fits to the results of our calculations
(Fig. \ref{fig: qdefs}). For each 
output size distribution at time $t_i$, we construct cumulative size distributions 
$n_{c}(r)$ for sets of eight annuli centered at distances $a_j$ from the central star. 
With $N$ = 64 annuli in our calculations, we have eight cumulative size distributions at each 
timestep.  We adopt a cumulative size distribution in the form of a power law 
$n_c \propto r^{-q}$ and derive the best-fitting exponents for planetesimals with 
radii $r$ = 0.1--1~km (\qone), $r$ = 1--10~km (\qtwo), $r$ = 10--100~km (\qthree), and 
$r \ge$ 100~km (\qfour).  To minimize fluctuations due to Poisson statistics, we restrict 
these fits to bins with $n_{c}(r) \ge 10 $.

Although we could fit the size distribution for $r \gtrsim$ 1~km to the product of a power 
law and an exponential \citep[e.g.,][]{weth1990,tanaka1994,lee2000},  several factors 
influence our choice of piece-wise fits to the overall size distribution. In our calculations,
we set the initial maximum sizes of planetesismals at $r_0$ = 1, 10, and 100~km. Measuring 
slopes between these sizes allows us to understand how our choices impact our results.
Mergers and 
destructive collisions often produce multiple inflection points in the size distribution
\citep[e.g.,][]{kb2004c,fraser2009b}. Fitting several distinct power laws allows us to identify 
the inflection points easily and to compare power laws at sizes smaller and larger than the 
inflection points.  Observations of TNOs are often well-fit by double power laws with a 
`break' at $r \approx$ 10--100 km 
\citep[e.g.,][]{fuentes2008,fuentes2009,fraser2009,fraser2010b,sheppard2010}.
Our $q_3$ and $q_4$ exponents allow us to compare calculations directly to observations
in the outer solar system. 

In the next sections, we consider the time-evolution of the four power law exponents and the size 
of the largest object in an annulus $r_{max}$. In every annulus, the growth of planetesimals
follows a similar path from orderly growth to runaway growth to oligarchic growth and the 
collisional cascade.  Thus, $r_{max}$ serves as a proxy for the evolution time $t$ and allows us
to compare results in different annuli at similar evolutionary states \citep[e.g.,][]{kb2008,kb2010}.
We begin with a discussion of the evolution of the size distribution at large sizes (\S3.1) and
then describe the evolution at smaller sizes in subsequent subsections.

\subsection{The Size Distribution of Large Objects} \label{sec: qfour}

\begin{figure}
\includegraphics[width=6.5in]{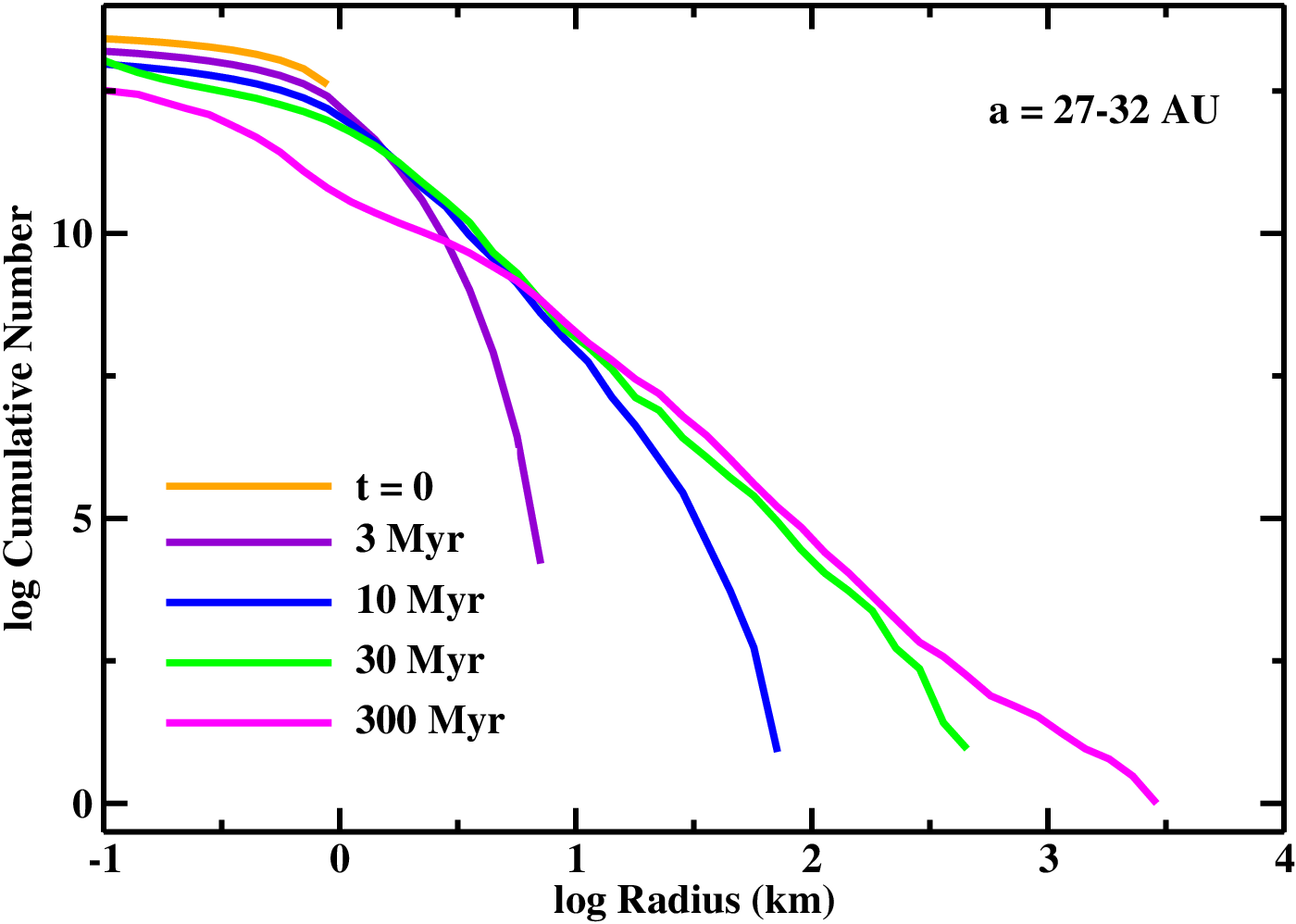}
\vskip 3ex
\caption{%
Time evolution of the cumulative number distribution for 8 annuli 
($a$ = 27--32~AU) in a disk with initial surface density distribution
$\Sigma = 0.18~x_m~(a / {\rm 30~AU})^{-3/2}$ g~cm$^{-2}$, $x_m$ = 1, 
and the $f_w$ fragmentation parameters.
The ensemble of planetesimals has an initial cumulative size distribution
$n_c \propto r^{-0.5}$ with a maximum radius of 1 km; thus, most of the 
initial mass is in the largest objects. During runaway growth at 1--10 Myr,
the largest protoplanets grow from $\sim$ 10~km to $\sim$ 1000~km. At the 
same time, the slope of the size distribution becomes shallower and shallower.
At late times ($t \gtrsim$ 100~Myr), the size distribution is a fairly
smooth power law from $\sim$ 3--10~km to $\sim$ 3000~km.
\label{fig: sd1}
}
\end{figure}

\begin{figure}
\includegraphics[width=6.5in]{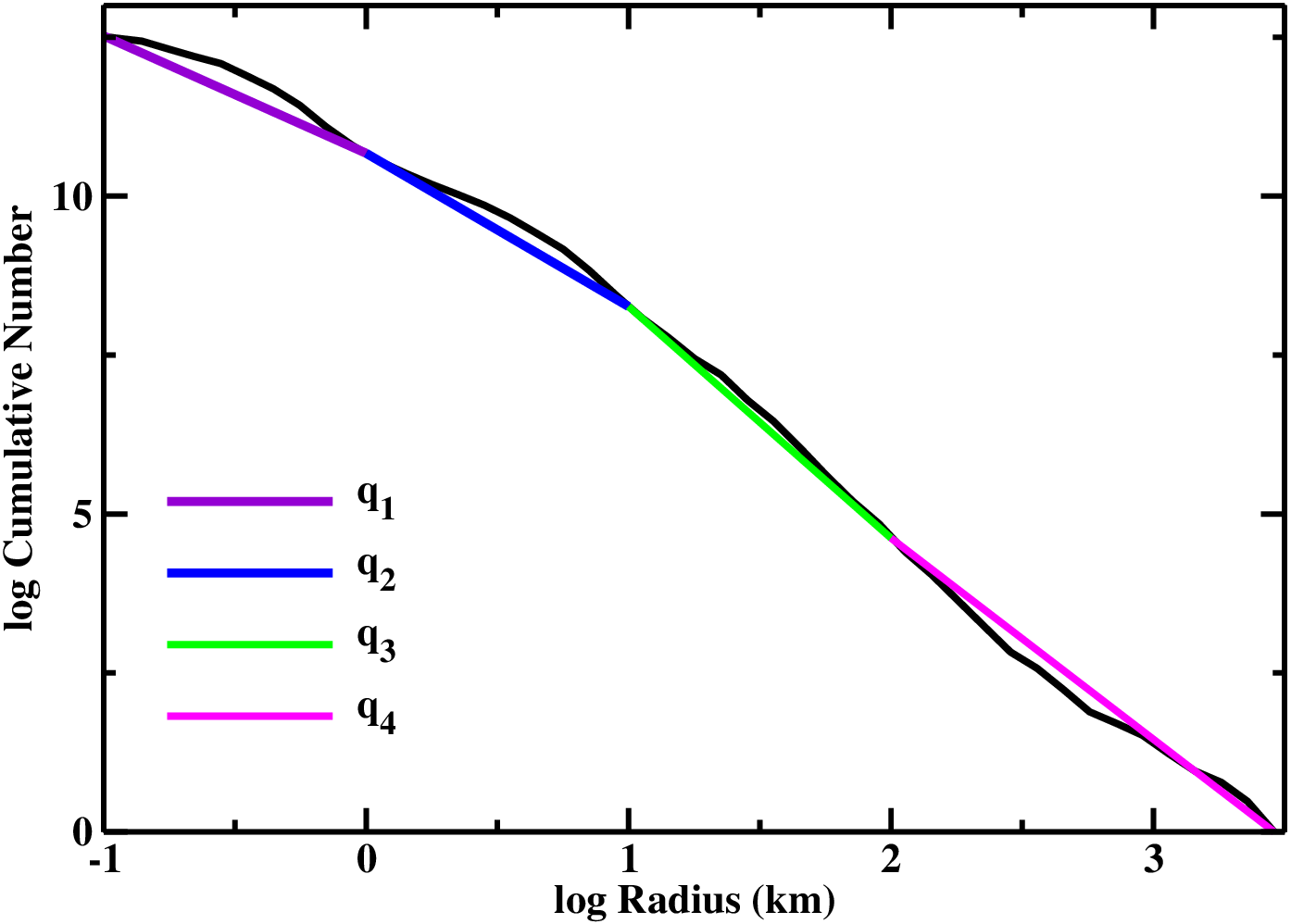}
\vskip 3ex
\caption{%
Definitions of the four slope parameters efined in the text. The black curve
repeats the 300~Myr size distribution from Fig. \ref{fig: sd1}.  The slopes 
\qone--\qfour are the slopes of the four line segments in the figure.
\label{fig: qdefs}
}
\end{figure}

\begin{figure}
\includegraphics[width=6.5in]{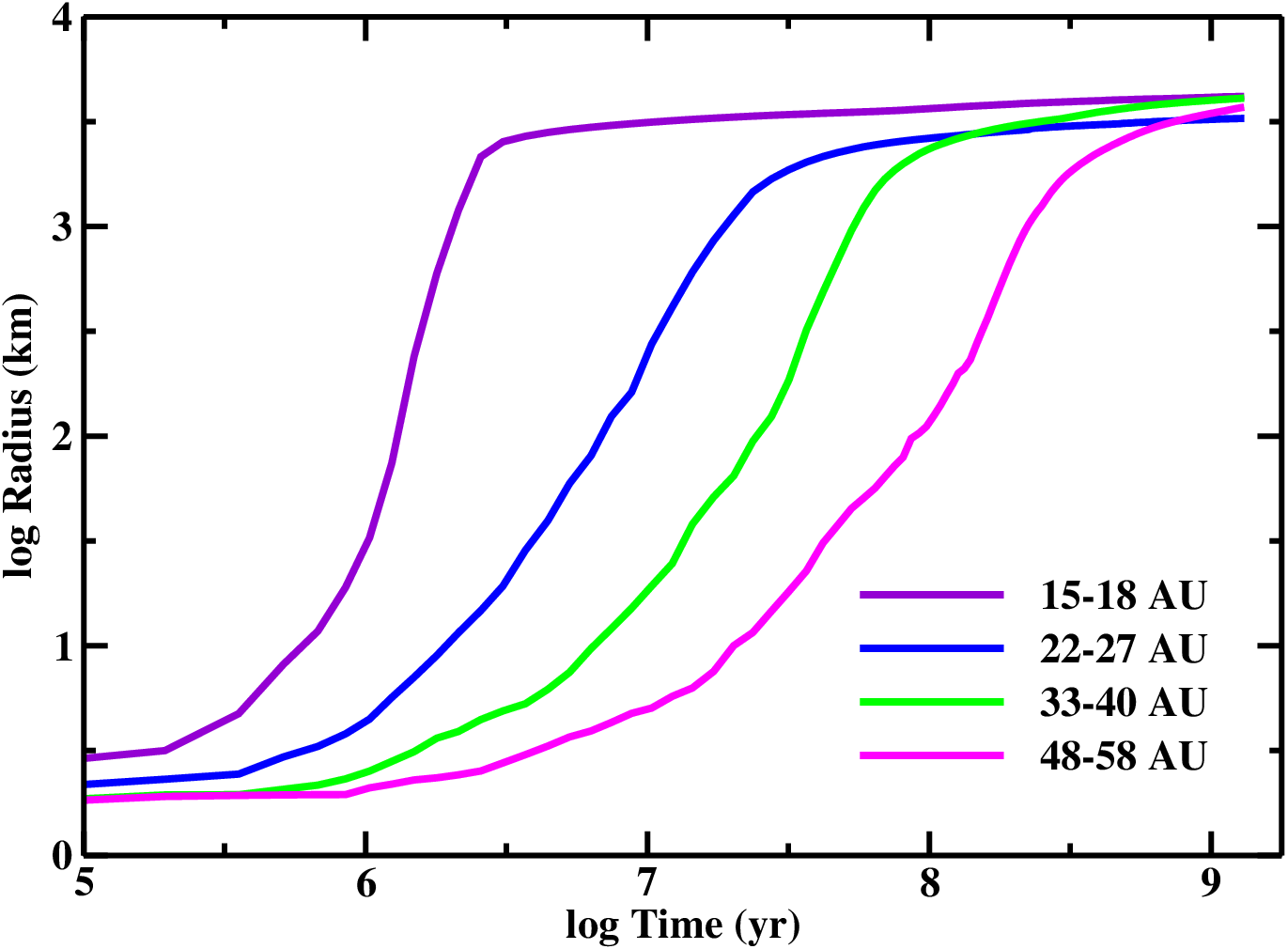}
\vskip 3ex
\caption{%
Time evolution of the size of the largest object in 4 sets of eight annuli
for the calculation described in Fig. \ref{fig: sd1}. After a short period
of slow growth, objects rapidly grow from $\sim$ 10~km to $\sim$ 1000~km
(runaway growth) and then grow more slowly to $\sim$ 3000--5000~km
(oligarchic growth). Objects close to the central star grow faster.  In 
roughly 1 Gyr, however, all objects reach a characteristic maximum radius 
of 3000--5000~km.
\label{fig: rmax1}
}
\end{figure}

Fig. \ref{fig: rmax1} illustrates the time evolution of $r_{max}$ for the calculation described
in Fig. \ref{fig: sd1}. In each annulus, the largest objects first grow slowly; $r_{max}$ grows
roughly linearly with time. Once runaway growth begins, large objects grow exponentially;
$r_{max}$ reaches $\sim$ 1000~km in 3~Myr at 15--18~AU, 20~Myr at 22--27~AU, 50~Myr at 33--40~AU, 
and 200~Myr at 48--58~AU. After planets achieve these radii, the collisional cascade removes
small planetesimals more rapidly than planets accrete them. Thus, planets grow much more slowly.
Despite the different timescales, planets in all annuli reach the same maximum size of roughly 3000~km.

\begin{figure}
\includegraphics[width=6.5in]{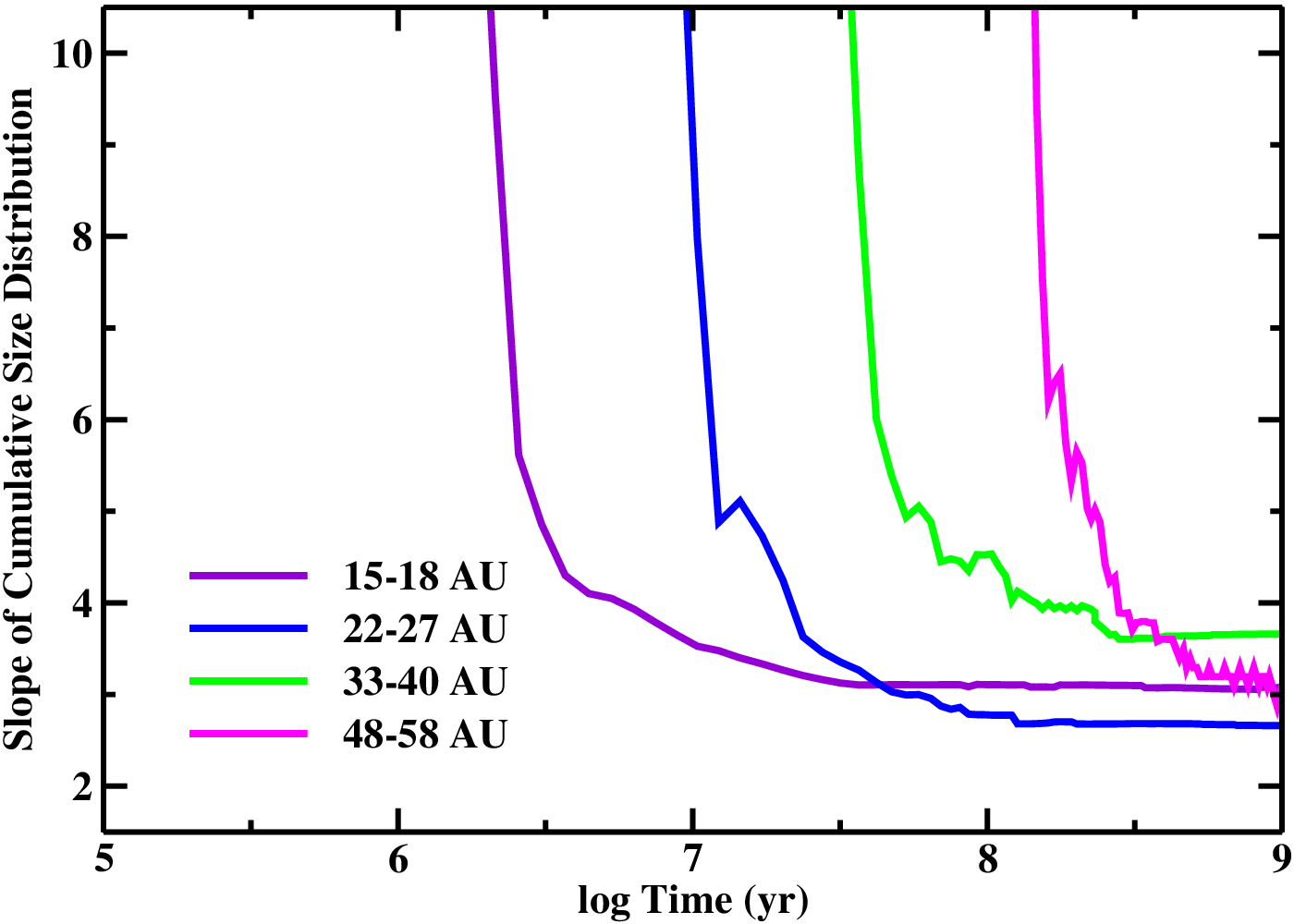}
\vskip 3ex
\caption{%
As in Fig. \ref{fig: rmax1} for the slope of the cumulative size distribution.
During runaway growth, the slope decreases rapidly from $q \gtrsim 10$ to
$q \approx 4$. As oligarchic growth proceeds, the slope asymptotically
approaches $q \approx 3$. Noise in the plots results from counting statistics 
in the mass bins.
\label{fig: slope1}
}
\end{figure}

Fig. \ref{fig: slope1} shows how \qfour\ evolves with time. When the first objects with $r > 100$~km
form, there are many 100~km objects and only a few with larger sizes. Thus, the size distribution at 
large sizes is very steep and \qfour\ is large. As the evolution proceeds, runaway growth produces 
more and more large objects from the ensemble of 100~km objects. With more objects at $r >$ 500~km and
fewer at 100~km, the power-law size distribution grows shallower and shallower (Fig. \ref{fig: sd1}); 
\qfour\ decreases.  In the late stages of 
oligarchic growth, planets reach their maximum radius. Because large objects mostly accrete material
from much smaller objects, the slope then converges on a characteristic value \qfour\ $\approx$ 3
\citep{gold2004,schlichting2011}.  Although the growth of the largest objects is monotonic in time 
(Fig. \ref{fig: rmax1}), small number statistics produces the fluctuations around the steady decrease 
of $q_4$ with time shown in the Figure.

To examine the relationship between \rmax\ and \qfour, we consider calculations for each set
of initial conditions ($x_m$, $r_0$, $q_{init}$, $f_i$). For each set of calculations, we collect 
all pairs (\rmax, \qfour) in all annuli at all times. We then bin these data in increments of
$0.1$ in log \rmax\ and calculate the median \qfour\ in each \rmax\ bin. This procedure yields 
the variation of median \qfour\ as a function of \rmax\ and the initial conditions. This median 
provides a good representation of the typical slope: the inter-quartile range is roughly 2 
when \rmax\ is small ($\sim$ 100~km) and roughly 0.5--1.0 when \rmax\ is large ($\gtrsim$ 1000~km).

\begin{figure}
\includegraphics[width=6.5in]{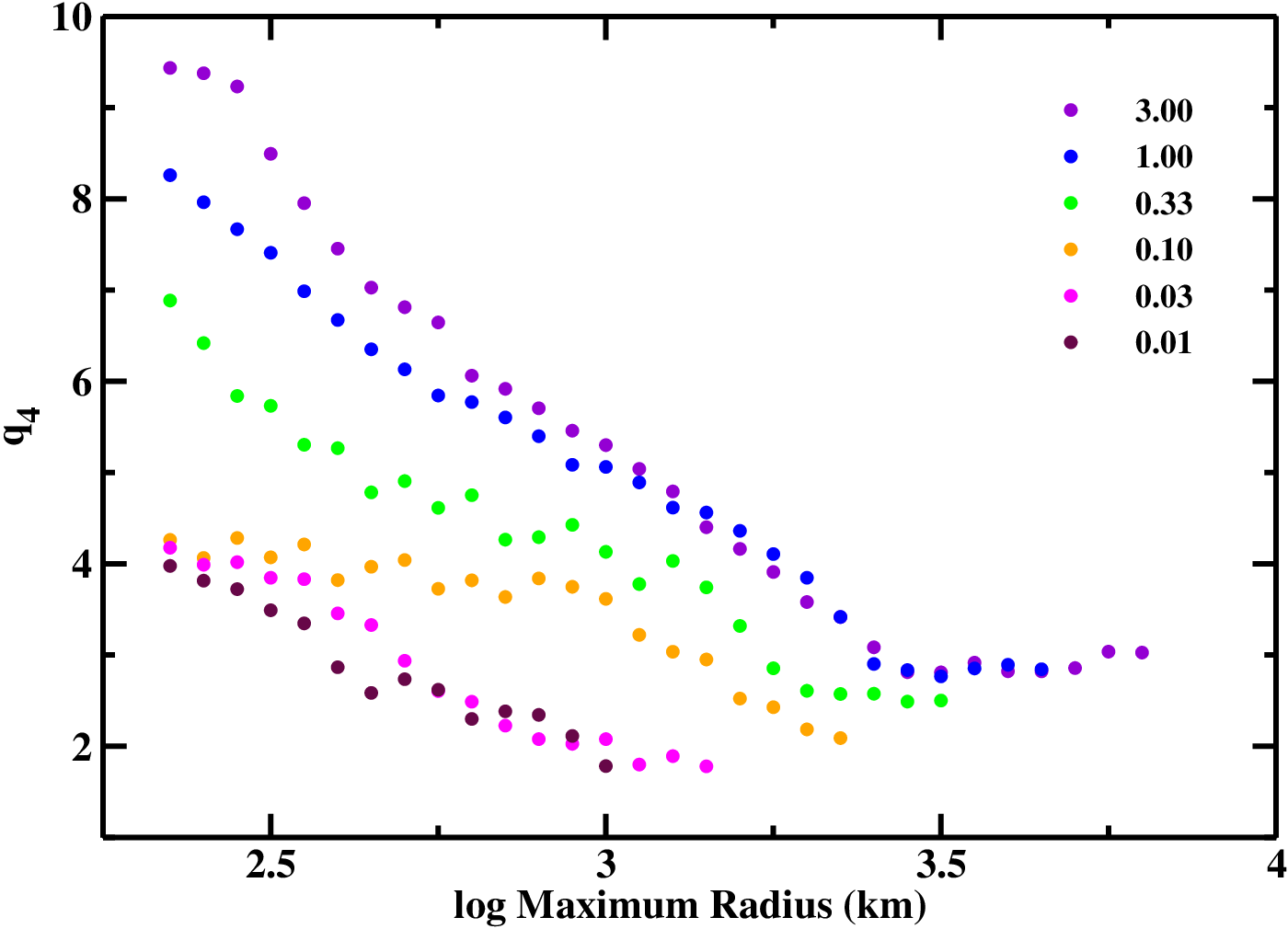}
\vskip 3ex
\caption{%
Variation of the median $q_4$ with $r_{max}$ for all annuli at all times in an 
ensemble of disks at 15--75~AU starting with most of the mass in the largest
planetesimals and the $f_w$ fragmentation parameters. The legend indicates $x_m$, 
the scaling factor for the initial surface density (eq. [4]).  The typical inter-quartile
range is roughly $\delta q_{iqr} \approx$ 2 at log $r_{max} \approx$ 2.25--2.75
and $\delta q_{iqr} \approx$ 0.5--1.0 at log $r_{max} \gtrsim$ 3.  For all $x_m$, the 
absolute lower bound in the slope is $q_4 \approx$ 2. The upper bound has $q_4$ 
roughly inversely proportional to $r_{max}$ and slowly decreasing with $x_m$.
In calculations with smaller $x_m$, slower growth allows the slope to reach the 
lower bound when the largest objects are relatively small.
\label{fig: rmax-slope1}
}
\end{figure}

Fig. \ref{fig: rmax-slope1} shows results for the complete set of calculations at $a$ = 15--75~AU
with $r_0$ = 1~km and the $f_w$ fragmentation parameters. For the ensemble of massive disks with
$x_m$ = 3 (violet points in the Figure), the median slope is large, \qfour\ $\approx$ 10, when 
\rmax\ $\approx$ 100--300~km.  For \rmax\ $\approx$ 200--2000~km, the median \qfour\ declines roughly 
monotonically with increasing \rmax. As \rmax\ grows from $\sim$ 2000~km to $\sim$ 5000~km, the 
median \qfour\ is roughly constant. 

The variation of \qfour\ with \rmax\ is sensitive to the initial disk mass. As the initial disk
mass declines from $x_m$ = 3 to $x_m$ = 0.01, growth times are longer and longer. Due to the
impact of gas drag and long-range stirring, small planetesimals in lower mass disks tend to
have larger eccentricities at the onset of runaway growth than those in higher mass disks.
Larger eccentricities reduce gravitational focusing factors, slowing growth and enabling 
stirring to grow faster. Because slower growth is more orderly, large objects have shallower 
size distributions.  When \rmax\ $\approx$ 100--300~km, lower mass disks thus have a smaller 
median \qfour\ than higher mass disks. Because growth is very slow for the smallest disk masses, 
all calculations with $x_m \lesssim$ 0.1 have roughly the same median \qfour\ $\approx$ 4. 

For disks with $x_m \gtrsim$ 0.3, gas drag (long-range stirring) is more (less) effective.
With larger gravitational focusing factors in more massive disks, growth is relatively fast. 
Once a few 100~km objects form, a stronger runaway leads to a more pronounced decline of
\qfour\ with \rmax.  As \rmax\ increases from roughly 100~km to $\sim$ 2000~km, the median 
\qfour\ declines roughly linearly with \rmax. Once \rmax\ $\gtrsim$ 2000~km, the median
\qfour\ is roughly constant.  In lower mass disks, the median \qfour\ declines more slowly 
with increasing \rmax. Because low mass disks never produce objects with \rmax\ $\gtrsim$ 
2000~km, the median \qfour\ is never constant with increasing \rmax.

\begin{figure}
\includegraphics[width=6.5in]{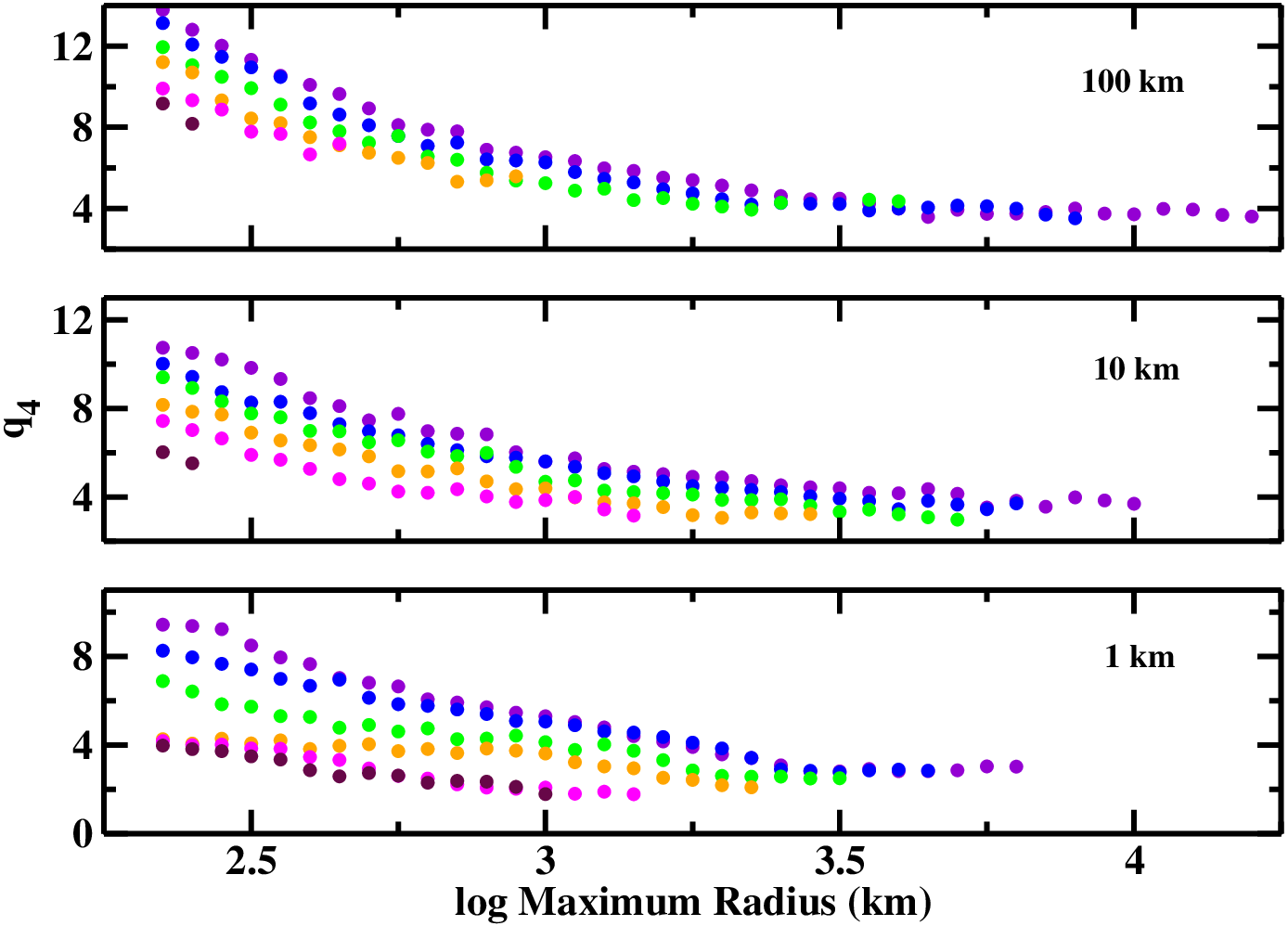}
\vskip 3ex
\caption{%
As in Fig. \ref{fig: rmax-slope1} for calculations with larger planetesimals.
The lower panel repeats Fig. \ref{fig: rmax-slope1}. The legend indicates
the initial radius of the largest planetesimals. Compared to calculations with
1~km planetesimals, calculations with 10--100~km planetesimals yield
larger planets and larger $q_4$ at fixed log $r_{max}$.
\label{fig: allrmax-slope1}
}
\end{figure}

In addition to the clear dependence on $x_m$, the median \qfour\ is also sensitive to 
the initial maximum planetesimal size $r_0$ 
(Fig. \ref{fig: allrmax-slope1}). For \rmax\ $\approx$ 100--300~km and $x_m \approx$ 1--3, 
the median \qfour\ steadily increases with larger $r_0$, from \qfour\ $\approx$ 8--9 for 
$r_0$ = 1~km to \qfour\ $\approx$ 10--11 for $r_0$\ = 10~km to \qfour\ $\approx$ 13 for 
$r_0$ = 100~km. Independent of $r_0$, the median \qfour\ declines steadily with larger 
\rmax. At the largest \rmax\ attained in our calculations, the median \qfour\ converges 
on an equilibrium slope $\sim$ 3--4. Calculations with smaller $r_0$ yield smaller values 
for this equilibrium slope.

\begin{figure}
\includegraphics[width=6.5in]{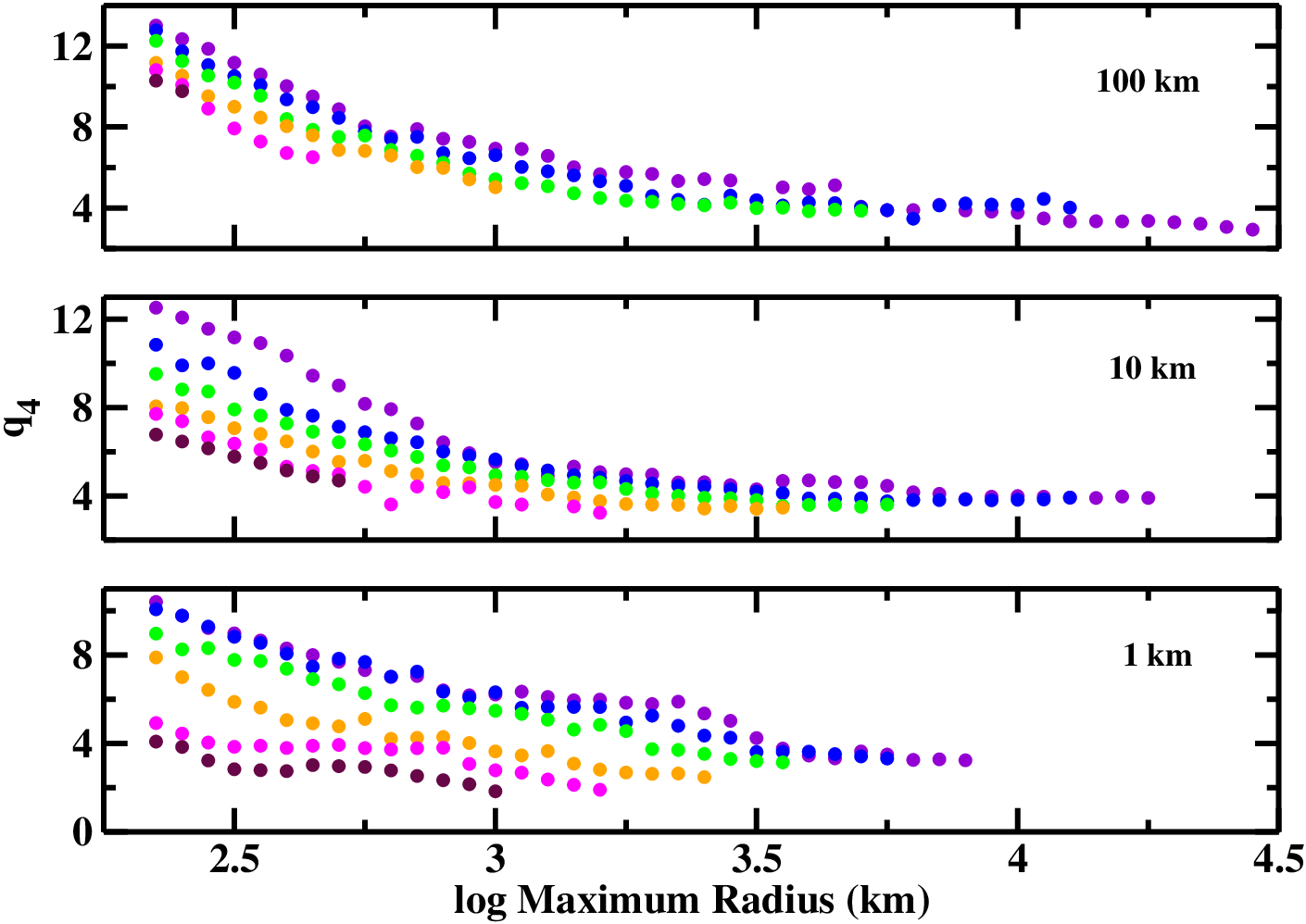}
\vskip 3ex
\caption{%
As in Fig. \ref{fig: allrmax-slope1} for disks at 30--150 AU over a 10~Gyr evolution
time.  Results at 30--150 AU yield the same relationship for $q_4$ as a function 
of $r_{max}$ and $x_m$. Longer evolution times in calculations at 30--150 AU produce 
larger \rmax\ than the calculations at 15--75 AU.
\label{fig: allrmax-slope2}
}
\end{figure}

The behavior of the \qfour--\rmax\ relation is remarkably independent of semimajor axis 
(Table \ref{tab: q4rmax}, Fig. \ref{fig: allrmax-slope2}).  In calculations at 30--150 AU, 
our 10~Gyr evolution times yield larger \rmax\ than the 1~Gyr evolution times for 
calculations at 15--75~AU. Otherwise, the median slope at small \rmax, the monotonic decline 
in slope at \rmax\ $\approx$ 300--2000~km, and the roughly constant slope for \rmax\ $\gtrsim$ 
2000~km closely follow the relations established at 15--75 AU. Our results show some 
indication for a smaller inter-quartile range in the median \qfour\ at larger semimajor 
axis. Confirming this trend requires a larger ensemble of calculations.

\begin{figure}
\includegraphics[width=6.5in]{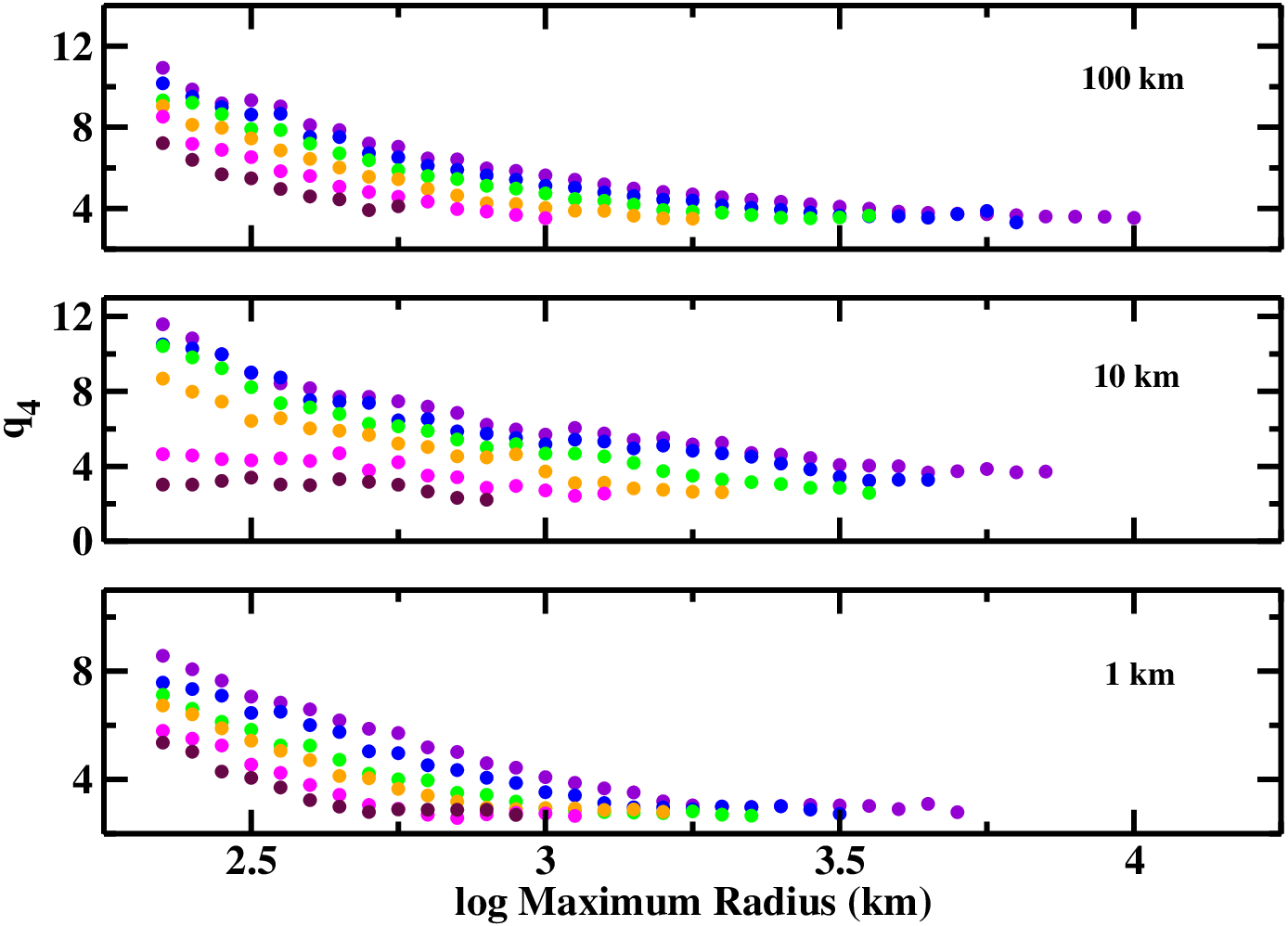}
\vskip 3ex
\caption{%
As in Fig. \ref{fig: allrmax-slope2} for disks at 30--150 AU starting with a mix of 
planetesimal sizes and the $f_w$ fragmentation parameters.  Calculations with a 
broader initial size distribution yield less massive planets and a shallower
relation between \qfour\ and \rmax.  For a fixed $r_{max}$, calculations with larger 
$r_0$ have steeper size distributions.
\label{fig: allrmax-slope3}
}
\end{figure}

Despite the insensitivity to semimajor axis, the median slope depends strongly on the initial
size distribution (Table \ref{tab: q4rmax}, Fig. \ref{fig: allrmax-slope3}). In calculations where
the initial mass is distributed among planetesimals with sizes of 1~m to 1~km, the median 
\qfour\ is systematically smaller ($\sim$ 1--2) throughout the evolution. This behavior is
independent of $r_0$. All of these calculations begin with a substantial amount of mass in
objects with sizes of 0.1~km or smaller. During runaway growth, collisional grinding removes
some of this material from the grid. Compared to calculations where all of the initial mass
is concentrated in the largest planetesimals, the largest objects in these calculations accrete 
material from a lower mass reservoir of small planetesimals which are easier to erode. Thus, 
the growth of the large objects is relatively slower and the median slope is smaller. Because 
calculations with larger $r_0$ have a smaller fraction of the initial mass in small planetesimals, 
this difference is smaller in calculations with larger $r_0$.

\begin{figure}
\includegraphics[width=6.5in]{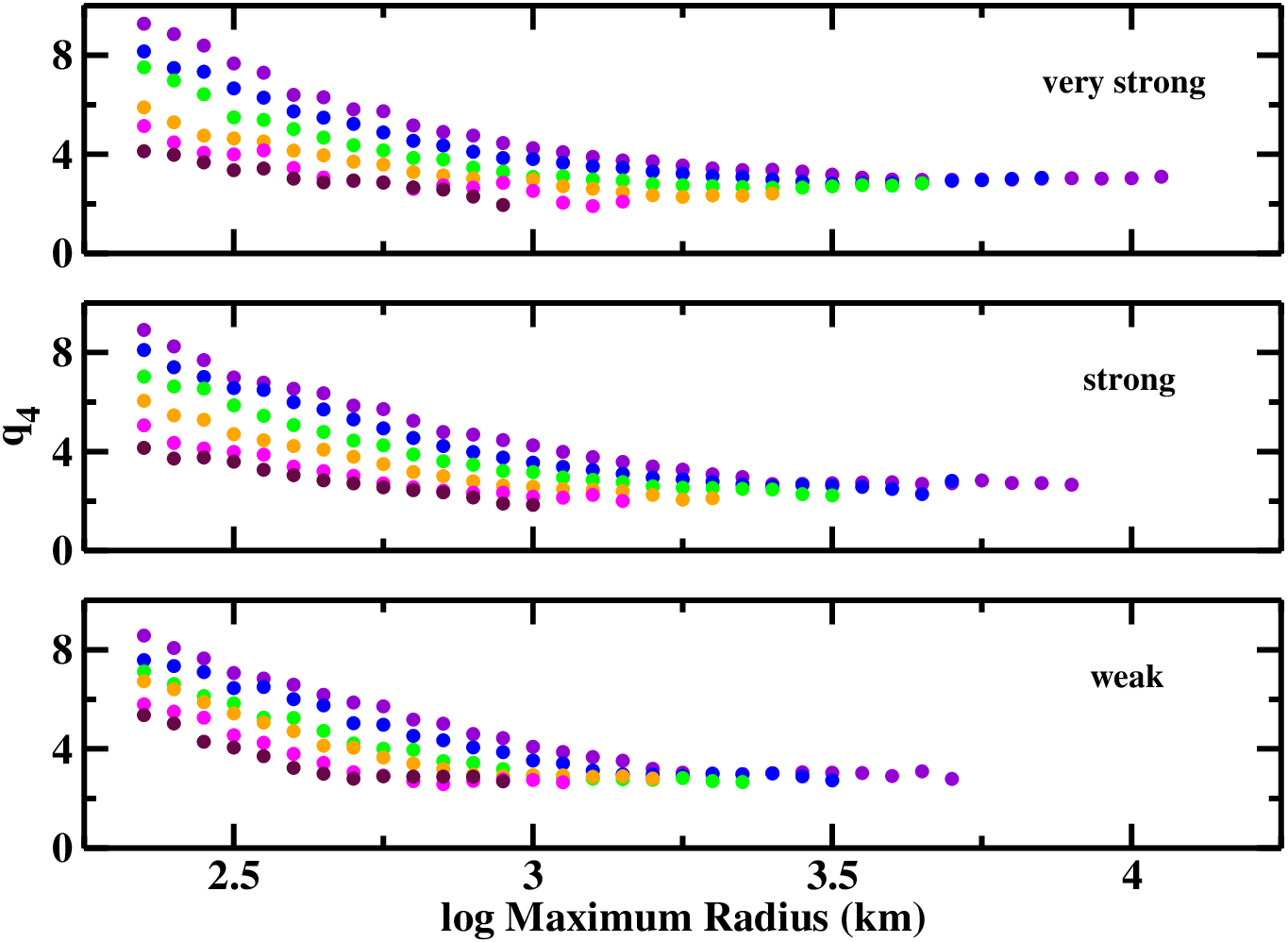}
\vskip 3ex
\caption{%
As in Fig. \ref{fig: allrmax-slope3} for calculations at 30--150 AU with the $f_w$ 
(lower panel), $f_s$ (middle panel), and $f_{vs}$ (upper panel) fragmentation parameters.  
All calculations begin with a broad initial size distribution of planetesimals and
$r_0$ = 1~km. Calculations with weaker planetesimals yield less massive planets.
\label{fig: allrmax-slope4}
}
\end{figure}

To conclude this discussion, Fig. \ref{fig: allrmax-slope4} shows how the \qfour--\rmax\ relation 
depends on the fragmentation parameters. For all $x_m$, larger planets grow in disks composed of 
stronger planetesimals.  Once the collisional cascade begins, collisions remove more mass from 
weak planetesimals than from strong planetesimals.  At identical evolution times, disks composed 
of weak planetesimals therefore have less mass in bins with small planetesimals. With oligarchs
accreting most of their mass from small planetesimals, they do not grow as rapidly or as large
when planetesimals are weak.

Despite the differences in \rmax, the median \qfour\ is fairly independent of the fragmentation
parameters (Table \ref{tab: q4rmax}). In the most massive disks with $x_m$ = 0.33--3, calculations 
with stronger planetesimals have slightly larger \qfour\ than calculations with weak planetesimals: 
\qfour\ = 2.9--4.1 for $f_w$, \qfour\ = 3.2--4.3 for $f_s$, and \qfour\ = 3.1--4.2 for $f_{vs}$.
These differences are smaller than the typical inter-quartile range of 0.5--1.0 when \rmax\ $\approx$
1000~km. In lower mass disks with $x_m \lesssim$ 0.1, calculations with stronger planetesimals tend 
to have shallower slopes, but the inter-quartile ranges are still much larger than the differences.

To summarize, Figs. \ref{fig: rmax-slope1}--\ref{fig: allrmax-slope4} demonstrate a clear relation
between the size of the largest object and the slope of the cumulative size distribution for
$r \gtrsim$ 100~km. For a broad range of initial conditions and fragmentation parameters, 
\qfour\ is large (6--12) when \rmax\ $\approx$ 300--500~km and small (2--5) when \rmax\ $\approx$ 
1000--3000~km. For any \rmax, disks with smaller masses have smaller \qfour\ than more massive
disks. 

To understand how the the rest of the size distribution evolves with \rmax, we now consider the
other three power-law exponents. We begin with the size distribution of 0.1--1~km objects in
\S\ref{sec: qone} and then discuss the evolution of 1--10~km (10--100~km) objects in 
\S\ref{sec: qtwo} (\S\ref{sec: qthree}).

\subsection{The Size Distribution of 0.1--1~km Objects}\label{sec: qone}

The slope of the cumulative size distribution for 0.1--1~km objects, \qone, is a proxy for the evolution 
of objects smaller than $r_0$. As the largest objects grow, they remove planetesimals from this size range. 
During runaway growth and the early stages of oligarchic growth, oligarchs accrete 
only a small fraction of the mass in 0.1--1~km objects (KB08, KB10).  Thus, the slope of this 
size distribution (\qone) should remain roughly constant with increasing \rmax. Once the collisional 
cascade begins, destructive collisions among small objects remove material from this range of sizes; 
destructive collisions among larger objects add material. Although the evolution of the slope then 
depends on the relative importance of destructive collisions among large and small objects, the 
collisional cascade eventually removes all of the mass in 0.1--1~km objects from every annulus 
(KB08, KB10). Thus, \qone\ should evolve dramatically during the cascade.

\begin{figure}
\includegraphics[width=6.5in]{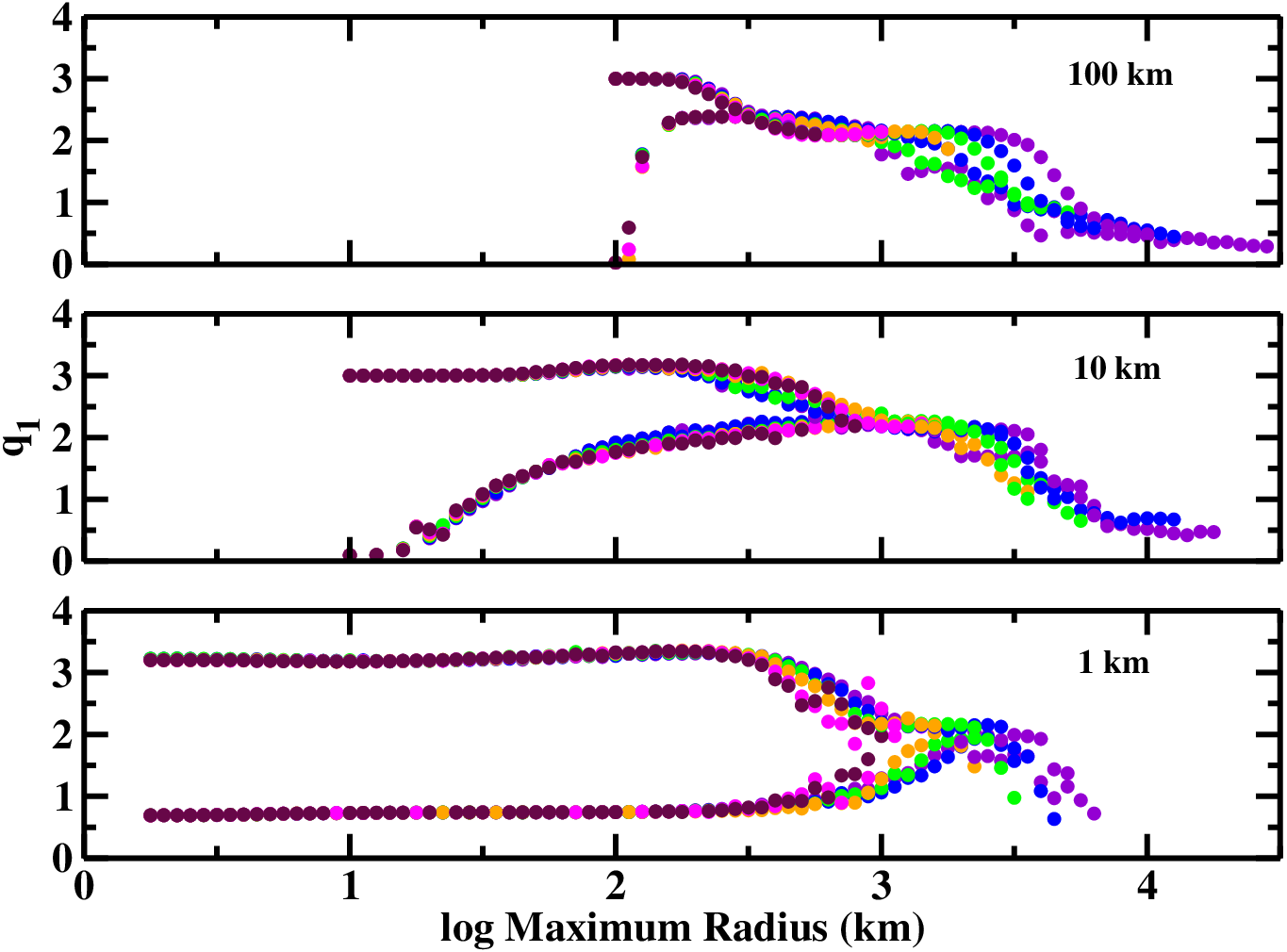}
\vskip 3ex
\caption{%
Variation of the median \qone\ with \rmax\ for calculations of disks at 30--150 AU, 
the $f_w$ fragmentation parameters, and $r_0$ = 1~km (lower panel), 10~km (middle
panel), and 100~km (upper panel). Results for calculations with $n_c \propto r^{-3}$
(upper branch of points in each panel) and $n_c \propto r^{-0.5}$ (lower branch of 
points in each panel) are included in all panels. Colors indicate results for different 
values of $x_m$ as in Fig. \ref{fig: rmax-slope1}.  The evolution of the slope is 
remarkably independent of $r_0$.  Until the onset of the collisional cascade at 
log \rmax\ $\approx$ 2.5--3, calculations `remember' the initial slope of the size 
distribution at 0.1--1~km. Once the collisional cascade begins, the slope rapidly 
evolves to \qone\ $\approx$ 2. As \rmax\ reaches $\sim$ 3000~km, the slope declines
to \qone\ $\approx$ 0.0--0.5.
\label{fig: q1rmax}
}
\end{figure}

Fig. \ref{fig: q1rmax} shows the variation of the median \qone\ with \rmax\ for all calculations 
with the $f_w$ fragmentation parameters. Each panel illustrates results for a different value of 
$r_0$; the lower (upper) set of points corresponds to calculations with initial $n_c \propto r^{-0.5}$ 
($n_c \propto r^{-3}$). Colors indicate initial disk mass ranging from \xm\ = 3 (violet) to
\xm\ = 0.01 (maroon) as in Fig. \ref{fig: rmax-slope1}. The remarkable overlap of colored points 
demonstrates that the evolution of the median \qone\ is independent of initial disk mass. 

For calculations with $r_0$ = 1~km, there are two main features of the evolution of \qone\ with 
\rmax.  As long as \rmax\ $\lesssim$ 300--500 km, \qone\ is constant. In the lower panel of 
Fig. \ref{fig: q1rmax}, \qone\ remains fixed at its initial value as the largest objects grow 
from sizes of a few km up to roughly 300~km. Thus, the size distribution of 0.1--1~km objects 
`remembers' the initial size distribution throughout runaway and oligarchic growth.

Once the collisional cascade begins, the general evolution of \qone\ is nearly independent of 
the initial size distribution and the initial size of the largest planetesimal.  As the largest 
objects grow to sizes of $\sim$ 1000~km, all size distributions begin to converge on 
\qone\ $\approx$ 2.0--2.5 (Table 3). After maintaining this limit for a small range in \rmax, 
\qone\ approaches a limiting value of 0.0--0.5. As the cascade removes nearly all of the mass
from an annulus, \qone\ remains at this limiting value.

The intermediate value of \qone\ $\approx$ 2.0--2.5 results from collisional erosion (see the Appendix
for a simple derivation). When the cascade begins, the ratio of the collision energy to the 
binding energy is $Q_c/Q_D^* \approx$ 0.1--1. In this regime, collisions produce a modest net 
reduction in the mass of the larger object. Together with the mass of the smaller object, this 
material is lost as debris and accumulates in much smaller mass bins.  Smaller objects are easier 
to erode. Thus, the size distribution evolves from the growth limit of roughly 3--4 to an 
intermediate limit of roughly 2.0--2.5. 

The final limiting value of \qone\ results from complete shattering (see the Appendix). As the 
cascade continues, the largest objects grow. Collision velocities among the smaller objects also 
grow. Once $Q_c/Q_D^* \gtrsim$ 2--3, nearly all of the mass of two colliding objects goes into 
debris.  Within the 0.1--1~km size range, all objects are susceptible to shattering. When 
shattering of the small objects becomes important, \qone\ approaches zero \citep{kb2004c}.

For calculations with $r_0$ = 10--100~km, 0.1--1~km objects collide and merge into larger objects 
before the collisional cascade begins.  Based on the evolution of the largest objects in
Figs. \ref{fig: rmax-slope1}--\ref{fig: allrmax-slope4}, growth tends to produce size distributions
with $q \approx$ 2--4. Thus, when the initial size distribution is steep ($n_c \propto r^{-3}$),
there is little evolution in \qone\ for \rmax\ $\lesssim$ 300--500~km (Fig. \ref{fig: q1rmax},
upper sets of points in the middle and upper panels). When the initial size distribution is
shallow ($n_c \propto r^{-0.5}$), slow growth leads to a gradual evolution from \qone\ $\approx$ 0
to \qone\ $\approx$ 2--3 (Fig. \ref{fig: q1rmax}, lower sets of points in the middle and upper panels).

Aside from this difference in the early evolution of \qone, these calculations also approach the 
two limiting values earlier in their evolution (at smaller \rmax). When $r_0$ = 10--100~km, there is
initially less mass in the 0.1--1~km range; thus, the collisional cascade modifies \qone\ earlier. 
At the onset of the collisional cascade, calculations with $r_0$ = 1~km have most of their mass in 
0.1--1~km planetesimals. Because it takes more collisional debris to change \qone\ in these 
calculations, \qone\ begins to change when \rmax\ is larger. In these calculations, \qone\ also 
has a larger dispersion as a function of initial disk mass and \rmax.

Analyzing calculations with other fragmentation parameters yields nearly identical results 
(Table 3). The trend of the evolution is always independent of initial disk mass. 
For $r_0$ = 1~km, \qone\ always remains close to its initial value. When $r_0$\ = 10--100~km, 
\qone\ remains roughly constant (when the initial \qone\ is larger than 2) or slowly evolves to
\qone\ $\approx$ 2--3 (when the initial \qone\ is smaller than 2). After the cascade commences, 
\qone\ first evolves to the canonical slope for erosive collisions of 2.0--2.5 (at \rmax\ $\approx$ 
1000~km) and then to the canonical shattering slope of roughly 0.0--0.5 (at \rmax\ $\approx$ 
3000--5000~km).  Although it always occurs at \rmax\ $\approx$ 500--1500~km, the transition from 
the initial \qone\ to the canonical \qone\ depends on the amount of mass initially in 0.1--1~km 
objects.  Disks with less (more) mass in these objects make the transition at smaller (larger) 
\rmax\ and earlier (later) in evolution time.

\subsection{The Size Distribution of 1--10~km Objects}\label{sec: qtwo}

Depending on the initial size of the largest object, the slope of the cumulative size distribution 
for 1--10~km objects, \qtwo, serves two distinct roles.  For calculations with $r_0$ = 1~km, 
the median \qtwo\ first measures changes in the slope as objects grow to sizes of 10~km and then tracks the
changes in the size distribution of these objects throughout the rest of runaway/oligarchic growth 
and the collisional cascade. In these models, we expect the median \qtwo\ first to decline with increasing 
\rmax\ (as in the evolution of \qfour\ with \rmax) and then to evolve to a slope characteristic 
of the collisional cascade.  For calculations with larger objects, $r_0$ = 10--100~km, changes in 
the size distribution for 1--10~km objects should follow the changes in \qone\ observed for the 0.1--1~km 
objects: the median \qtwo\ should evolve slowly to $\sim$ 2--3 during runaway/oligarchic growth and then 
evolve more rapidly to a smaller value during the collisional cascade.

\begin{figure}
\includegraphics[width=6.5in]{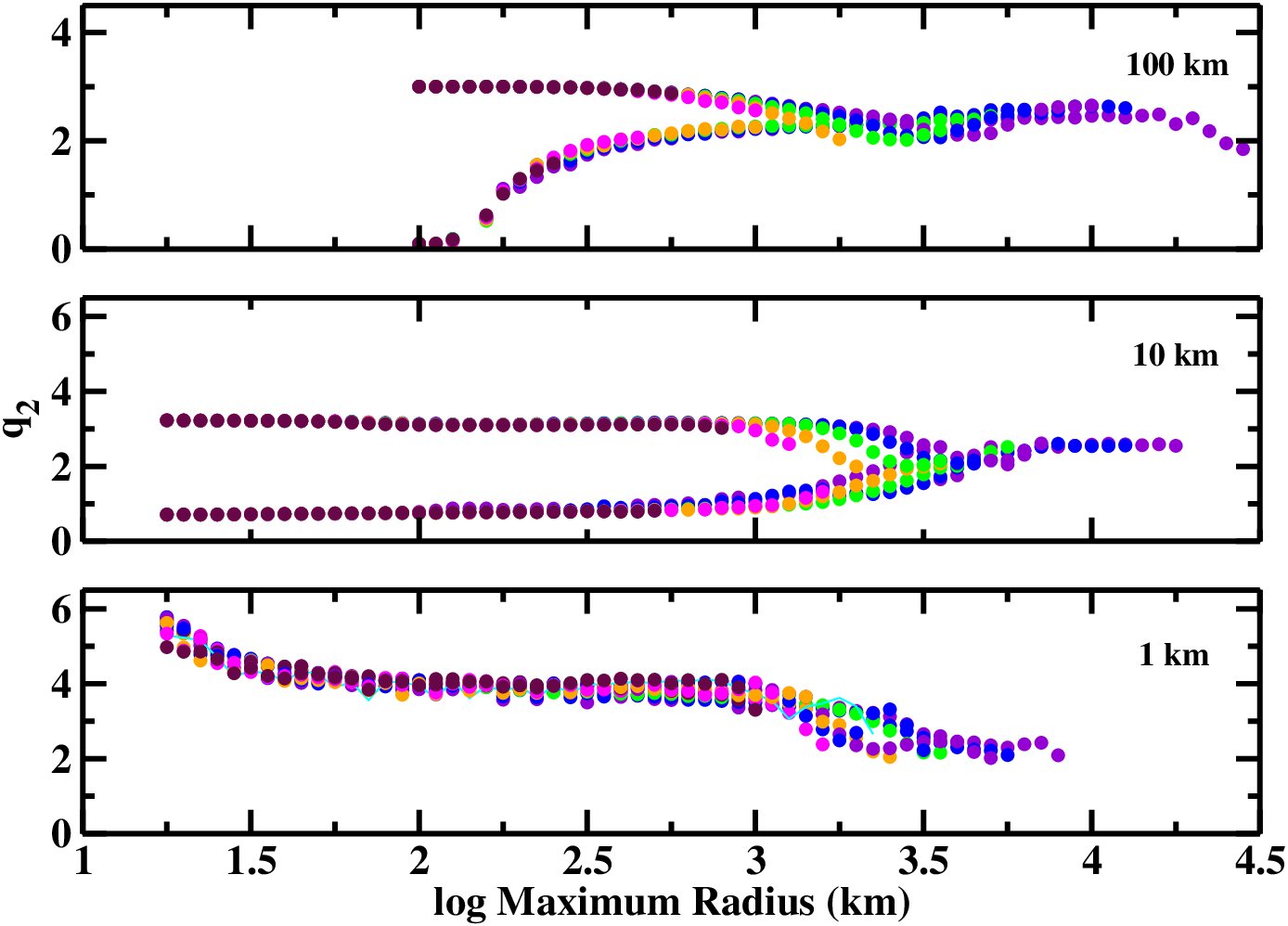}
\vskip 3ex
\caption{%
As in Fig. \ref{fig: q1rmax} for \qtwo, the slope of the size distribution at 1--10~km.
For calculations with small $r_0$, the evolution of \qtwo\ is independent of the slope of
the initial size distribution at smaller sizes; as \rmax\ grows from $\sim$ 30~km to
$\sim$ 1000~km, \qtwo\ monotonically declines from \qtwo\ $\approx$ 5--6 to \qtwo\ $\approx$ 4.  
When $r_0$ is larger, \qtwo\ remains close to its initial value until the onset of the 
collisional cascade at log \rmax\ $\approx$ 3. When the cascade begins, \qtwo\ evolves to 
$\sim$ 2 and remains close to this value independent of \rmax.
\label{fig: q2rmax}
}
\end{figure}

Fig. \ref{fig: q2rmax} shows the variation of the median \qtwo\ with \rmax\ for all calculations 
with the $f_w$ fragmentation parameters. The format is identical to Fig. \ref{fig: q1rmax},
with separate panels for each $r_0$, colors indicating the initial disk mass, and an upper
and lower set of points corresponding to our two adopted initial size distributions. As in
Fig. \ref{fig: q1rmax}, the evolution of the median \qtwo\ as a function of \rmax\ is 
independent of the initial disk mass.

In each panel, the evolution of \qtwo\ follows expectation. In the lower panel ($r_0$ = 1~km), 
the median \qtwo\ declines from $\sim$ 5--6 when \rmax\ $\approx$ 20~km to $\sim$ 4 for 
\rmax\ $\gtrsim$ 100~km. This evolution is independent of the initial size distribution.  Once 
the collisional cascade begins, \qtwo\ declines to $\sim$ 2.0--2.5 as \rmax\ grows from 1000~km
to 3000~km. At early times, the slope of the 1--10~km size distribution is similar to the slope 
of the large object size distribution (\qfour); at late times, the slope evolves into one of
the canonical slopes for the collisional cascade.

For larger $r_0$, the evolution of \qtwo\ tracks features of the evolution of \qone. In the middle 
panel of Fig. \ref{fig: q2rmax}, \qtwo\ follows much of the \qone\ evolution for 1~km objects 
(Fig. \ref{fig: q1rmax}, lower panel).  During runaway and oligarchic growth, \qtwo\ remains 
constant with \rmax. Once the cascade begins, \qtwo\ evolves to the canonical value for the 
early stages of the collisional cascade \qtwo\ $\approx$ 2.0--2.5. As \rmax\ approaches $10^4$ km, 
\qtwo\ remains constant at roughly 2.5 instead of dropping to $\sim$ 0.

In the upper panel, the evolution of \qtwo\ for $r_0$ = 100~km similarly tracks the \qone\ evolution 
for $r_0$ = 10~km.  As the largest objects grow from $\sim$ 100~km to $\sim$ 1000~km, 1--10~km
objects collide and merge into larger objects. Growth tends to evolve the slope to 2--4. For
calculations with $n_c \propto r^{-3}$ (upper set of points), \qtwo\ does not change. When the
initial size distribution is concentrated into large objects ($n_c \propto r^{-0.5}$), \qtwo\ slowly
approaches 2. Eventually, the collisional cascade begins; \qtwo\ evolves to 2--2.5 and remains
in this range as \rmax\ reaches roughly $10^4$ km. 

The ratio of the collision energy $Q_c$ to the binding energy $Q_D^*$ allows us to understand the
difference between the \qone\ evolution for $r_0$ = 1~km and the \qtwo\ evolution for $r_0$\ = 10--100~km.
When the collisional cascade begins, erosive collisions remove mass from the 0.1--1~km and the 
1--10~km size ranges.  During this evolution, both \qone\ and \qtwo\ evolve to 2.0--2.5. Once 
3000~km $\lesssim$ \rmax\ $\lesssim$ $10^4$ km, collisions among smaller planetesimals produce 
shattering. However, the larger planetesimals remain in the erosive regime.  As a result, objects 
are completely removed from the ensemble of 0.1--1~km objects (producing \qone\ = 0.0--0.5) and 
slowly eroded among the ensemble of 1--10~km objects (leaving \qtwo\ at 2.0--2.5). 

Results for calculations with the $f_s$ and $f_{vs}$ fragmentation parameters are nearly identical 
to those with the $f_w$ fragmentation parameters. When collisions produce mergers instead of debris, 
\qtwo\ evolves from its initial value to the canonical `merger' limit of 2--4. When $r_0$ is 1~km,
\qtwo\ has no memory of the initial size distribution. Calculations with larger $r_0$ remember the
initial \qtwo\ and maintain this slope throughout much of the evolution.  Once the cascade begins, 
\qtwo\ always evolves to $\sim$ 2.  Because 1--10~km objects are not completely shattered at any 
point in the collisional cascade, \qtwo\ always remains larger than $\sim$ 2 and never drops to 0--1.

\subsection{The Size Distribution of 10--100~km Objects}\label{sec: qthree}

Based on our results for \qone\ and \qtwo, the evolution of \qthree, the slope for 10--100~km objects,
should follow one of two paths. When the initial size of the largest planetesimal is 10~km or smaller,
planetesimals must grow to fill the 10--100~km size bins. Growth produces a characteristic evolution
of the slope from large values to $\sim$ 2--4 (Fig. \ref{fig: rmax-slope1}). As the evolution makes 
the transition from growth to the collisional cascade, the slope evolves to a canonical value of 
roughly 2. Because 10--100~km objects are harder to shatter than 1--10~km objects, the slope produced
by the cascade should never fall below 2. When planetesimals have initial sizes larger than 10 km, 
collisional growth requires a long time to erase the initial size distribution. Eventually, growth and 
the cascade modify the size distribution, which should approach the canonical value of 2.

\begin{figure}
\includegraphics[width=6.5in]{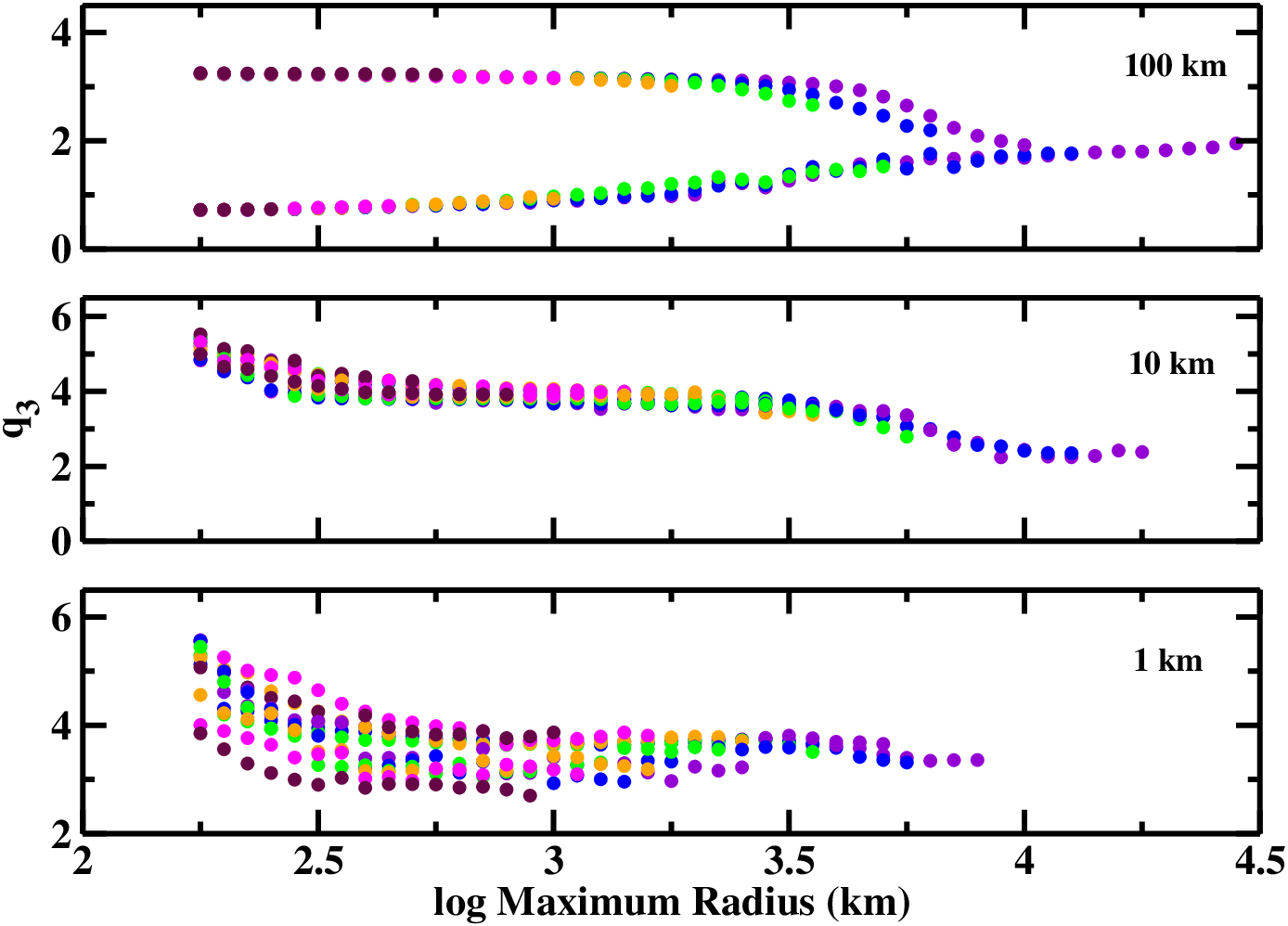}
\vskip 3ex
\caption{%
As in Fig. \ref{fig: q1rmax} for \qthree, the slope of the size distribution at 10--100~km.
For calculations with $r_0 \lesssim$ 10~km, the evolution of \qthree\ is independent of the 
slope of the initial size distribution at smaller sizes; as \rmax\ grows from $\sim$ 
300~km to $\sim$ 1000~km, \qthree\ monotonically declines from \qthree\ $\approx$ 5--6 to 
\qthree\ $\approx$ 3--4. At any log \rmax\ $\lesssim$ 3, the variation in \qthree\ with $x_m$
is larger for calculations with smaller $r_0$. Once the collisional cascade begins, the 
slope slowly evolves to \qthree\ $\approx$ 2--3, with \qthree\ $\approx$ 2 favored at late times.
\label{fig: q3rmax}
}
\end{figure}

Fig. \ref{fig: q3rmax} confirms these expectations for calculations with the $f_w$ fragmentation
parameters. Following the format of Figs.  \ref{fig: q1rmax}--\ref{fig: q2rmax}, there is a panel 
for each $r_0$, colors indicating the range in $x_m$, and sets of points for each initial size
distribution. Once again, the evolution is nearly independent of the initial disk mass.

For calculations with $r_0$ = 1~km, growth initially produces large \qthree\ $\approx$ 4--6. Shallower
initial size distributions and more massive disks tend to yield larger \qthree. As objects grow, all
calculations converge on a smaller range of \qthree, $\sim$ 2.5--3.5, with smaller initial disk
masses favoring smaller \qthree. For \rmax\ $\gtrsim$ 1000--2000~km, \qthree\ is independent of
initial disk mass.

When $r_0$ = 10~km, growth leads to a smaller range of initial median slopes, \qthree\ $\approx$ 5--6 
for \rmax\ $\approx$ 200~km. These median slopes are independent of the initial size distribution and 
the initial disk mass. As the largest objects grow to \rmax\ $\approx$ 1000~km, \qthree\ reaches an 
approximate equilibrium at $\sim$ 4. Once the cascade begins, this slope gradually declines to 
\qthree\ $\approx$ 2.

This evolution for \qthree\ follows the evolution of \qtwo\ for calculations with $r_0$ = 1~km.
Independent of the initial size distribution, growth produces a characteristic size distribution
which slowly evolves to shallower slopes as the largest objects grow. Eventually, the cascade
changes the size distribution and the median \qthree\ becomes smaller. In the middle panel of
Fig. \ref{fig: q3rmax}, this change occurs when \rmax\ $\gtrsim$ 3000~km. A similar, but less 
obvious, change in slope occurs for the $r_0$ = 1~km calculations in the lower panel.

When $r_0$ = 100~km, \qthree\ remembers the initial size distribution throughout most of the 
evolution (Fig. \ref{fig: q3rmax}, upper panel). In these calculations, the initial size 
distribution is unaffected by runaway/oligarchic growth and the early stages of the collisional 
cascade.  Throughout these phases of the evolution, growth is slow and collisions have little
impact on large objects. Thus, the size distribution is unchanged. During the late stages of the 
cascade, occasional collisions among large objects produce some debris which fills all mass bins in 
the 10--100~km size range. Independent of the initial size distribution, this debris drives the 
slope of the size distribution to the canonical value of \qthree\ = 2.0--2.5.

The evolution of \qthree\ is independent of the fragmentation parameters (Table 3). When the
planetesimals are initially small, \qthree\ first evolves from large values ($\sim$ 4--6) to
small values ($\sim$ 3). Once the cascade begins, \qthree\ evolves to 2.0--2.5. In this size 
range, the collisional cascade has less impact than at smaller sizes. Thus, the transition from 
the \qthree\ $\approx$ 3--4 characteristic of runaway/oligarchic growth to \qthree\ $\approx$ 
2.0--2.5 occurs at larger \rmax\ (and later evolution time).

\subsection{Summary of Size Distribution Evolution}
\label{sec: summ-results}

At 15--150~AU, the evolution of planetesimals follows a similar path at every distance from
the central star.  Initially, collisional damping and gas drag reduce the velocities of small
objects. Among larger objects, low velocity collisions produce mergers instead of debris.  Along 
with dynamical friction and viscous stirring, damping produces large gravitational focusing 
factors; this evolution leads to runaway growth of the largest planetesimals. As these protoplanets
grow, they stir up the leftover smaller planetesimals. Eventually, collisions among the leftovers
produce debris instead of mergers. A collisional cascade ensues, which grinds planetesimals with
sizes of 1--3~km and smaller to dust.

Despite the similarity in the evolutionary path, there is considerable diversity in the overall 
outcome. For identical starting conditions, the stochastic nature of coagulation leads to a broad 
range in the rate protoplanets grow, in the maximum radius of protoplanets, in the distribution of 
sizes as a function of time, and in the amount of mass lost through the collisional cascade. 
Although this diversity does not mask clear trends in the median outcome as a function of initial
conditions \citep{kb2008,kb2010}, the chaotic nature of the evolution prevents associating a single
outcome with a single set of starting conditions.

To quantify the overall evolution of the size distribution of icy planetesimals at 15--150~AU, 
we define four slopes, \qone--\qfour, which measure the best-fitting slope of a power-law 
size distribution between 0.1--1~km (\qone), 1--10~km (\qtwo), 10--100~km (\qthree), and 100~km 
to \rmax\ (\qfour). For all \rmax\ $\gtrsim$ 200~km, we derive the median slopes and the 
inter-quartile range as a function of \rmax.  We use median \qone--\qfour\ and the inter-quartile
range to track the evolution of the size distribution and to measure the range of outcomes as
a function of initial conditions.

Based on these slopes, we infer several general features of the evolution:

\begin{itemize}

\item For all distances from the central star, the growth of icy planets leads to size distributions 
with \q\ $\approx$ 2--4 \citep[see also][]{kl1999b,kenyon2002,gold2004,schlichting2011}. This size
distribution results from large objects accreting primarily small objects rather than other large 
objects. As largest objects grow, the slope evolves from \qfour\ $\approx$ 6--14 when \rmax\ $\approx$
200~km to \qfour\ $\approx$ 2--4 when \rmax\ $\gtrsim$ 1000~km. 

\item The collisional cascade produces shallower slopes in the size distribution, \q\ $\simless$ 2.0--2.5.
During the cascade, several processes set the size distribution for objects with $r \gtrsim$ 0.1~km
\citep[see also][]{obrien2003,kb2004c,obrien2005,pan2005}:
(i) growth by mergers (\q\ $\approx$ 3), 
(ii) debris production \citep[\q\ $\approx$ 2.5--3, e.g.,][]{dohn1969,obrien2003},
(iii) erosion (\q\ $\approx$ 2.0--2.5; see the Appendix)
and (iv) shattering \citep[\q\ $\approx$ 0.0--0.5, e.g.,][and the Appendix]{kb2004c}.  
At the smallest sizes ($r \lesssim$ 0.01--0.1~km), debris production dominates the size distribution 
and \q\ $\approx$ 3.5--4 \citep{obrien2003,kb2004c,kb2005,obrien2005,kriv2005,kb2008,benavidez2009,fraser2009b,kb2010}. 
At larger sizes, erosion slowly removes mass from all objects and produces \q\ $\approx$ 2.0--2.5. 
Once the collisional cascade reaches its peak (\rmax\ $\gtrsim$ 3000~km), 
shattering of 0.1--1~km objects leads to \q\ $\approx$ 0.0--1.0.

\item The evolution of the large object size distribution is independent of radial distance from the 
central star and is generally independent of the fragmentation parameters. However, the evolution of 
\qfour\ is sensitive to the initial size of the largest planetesimal ($r_0$) and the initial mass of 
the disk ($x_m$).  Lower mass disks evolve over a smaller range of \qfour. Calculations with larger 
planetesimals tend to produce larger median \qfour. 

\item For any \rmax, the range in \qfour\ is fairly large. At small sizes (\rmax\ $\approx$ 200--500~km),
the inter-quartile range about the median \qfour\ is roughly 2. This range drops to roughly 1 at
\rmax\ $\gtrsim$ 1000~km. For a typical \qfour\ $\approx$ 3, roughly 50\% of calculations with the
same initial conditions yield \qfour\ = 2.5--3.5. The rest have \qfour\ = 2--2.5 and \qfour\ = 3.5--5.

\item Among the intermediate mass objects with $r =$ 1--100~km, the evolution of the slope of the
size distribution depends on the initial size of the largest planetesimal.

\item When $r_0$ is small, growth produces objects with $r =$ 1--100~km. Growth erases knowledge of
the initial size distribution; the slope evolves to the standard `growth slope,' \qtwo\ $\approx$
\qthree\ $\approx$ 2--4. Once the collisional cascade begins, the slopes evolve to the canonical 
value of 2 for erosive collisions.

\item When $r_0$ is large, growth cannot change the initial slope of the size distribution. Thus,
\qtwo\ and \qthree\ remember the initial \q\ until the onset of the collisional cascade. Once
the cascade begins, erosive collisions lead to \qtwo\ $\approx$ \qthree\ $\approx$ 2.0--2.5.

\end{itemize}

\section{APPLICATION TO THE TRANSNEPTUNIAN REGION}
\label{sec: apps}

To apply these results to the formation of TNOs, we place our calculations in the context of recent 
observations of TNOs and models for the formation of the solar system. The observations allow us to 
set \rmax\ and the slopes of the size distribution in various size ranges. Observations of young 
stars in nearby star-forming regions and theories of solar system formation constrain the initial 
properties of the protosolar disk and the likely epoch(s) of TNO formation.  We begin with the 
theoretical issues and then summarize the direct observational constraints.

\subsection{Theoretical Constraints}
\label{sec: apps-theory}

Current theories for the origin of the solar system broadly explain the nature of TNOs
as a several step process.  During the early evolution of the protosolar disk, dust grains 
within the disk grow into icy planetesimals \citep[e.g.,][]{weiden2006,chiang2010} and then 
into icy protoplanets \citep[e.g.,][]{liss1987,kl1999a,kb2008,kb2010}.  Four protoplanets 
become gas giants \citep{pollack1996,ali2005,ida2008,LisHub09,bk2011a}.  Sometime after 
the gas giants reach roughly their final masses, dynamical interactions among the gas giants,
the much less massive icy protoplanets, and any leftover planetesimals result in the 
ejection or migration of icy objects into the transneptunian region and the Oort Cloud 
\citep[e.g.,][]{malhotra1993,hahn1999,morby2004b,tsig2005,gomes2005}.  Once dynamical 
evolution begins, lower surface densities and larger orbital velocities among the TNOs 
reduce growth rates and promote collisional erosion 
\citep[e.g.,][]{stcol1997b,kl1999a,kb2004c,obrien2005,kb2008,fraser2009b,kb2010}.  Thus, dynamical 
evolution ends the growth of TNOs.  

Despite the uncertainty in the theory, this picture constrains the initial mass of the protosolar
disk and the likely range of formation times for TNOs.  The protosolar disk probably has properties 
similar to the disks observed in nearby star-forming regions such as Ophiuchus, Orion, and 
Taurus-Auriga \citep[e.g.,][]{wil2011}. Infrared and radio observations demonstrate that these 
systems lose their gas and dust on timescales of 1--10~Myr 
\citep{haisch2001,currie2009,kk2009,wil2011}. For the solar system, these data constrain the initial 
disk mass, 0.001--0.1 \msun, and the timescale, 1--10~Myr, for the formation of gas giant planets.

Dynamical calculations of the solar system place additional limits on the initial disk mass and the
formation time.  Several results suggest some TNOs formed at 20--25 AU and then migrated outwards 
with the giant planets \citep[e.g.,][]{malhotra1993,malhotra1995,hahn1999,levison2003,MorbiSSBN}. 
To explain the orbital architecture of the giant planets, the TNOs, and the Oort cloud, these 
calculations rely on a massive disk of solids with $x_m \gtrsim$ 1 at 15--30~AU
\citep{hahn1999,levison2003,MorbiSSBN}.  Although these calculations place little direct constraint 
on the formation time, Figs. \ref{fig: rmax-slope1}--\ref{fig: allrmax-slope4} demonstrate that 
more massive disks have larger \qfour\ at fixed \rmax. Thus, relying on a massive disk to enable 
outward migration places indirect constraints on any combination of formation time, \rmax, and \qfour.

Various lines of evidence associate migrating gas giant planets with an increased rate of cratering 
during the Late Heavy Bombardment \citep[hereafter LHB; e.g.,][]{levison2001b,gomes2005,strom2005}. 
Although constraints on the cratering history prior to this epoch are weak \citep[e.g.,][]{chapman2007}, 
detailed analyses suggest the intense period of bombardment ended $\sim$ 500--700 Myr after the 
formation of the solar system \citep[e.g.,][]{tera1974,hartmann2000,stoffler2001}.  It is unlikely 
that the gas giants migrated after the LHB.  Thus, the end of the LHB sets an upper limit on the 
formation time for TNOs.

For our analysis, we concentrate on the formation of TNOs before the gas giants dynamically rearrange
the solar system. The initial conditions chosen for our calculations broadly match the range of
initial disk masses and semimajor axes applicable to TNO formation. By following the evolution of
TNOs for 1--10~Gyr, we cover the most likely range of formation times, $\sim$ 10--700~Myr. Thus, our
calculations can provide clear tests of the coagulation picture for TNO formation.

\subsection{Observational Constraints}
\label{sec: apps-obs}

Recent observations place excellent limits on the masses and radii of the largest TNOs.  All of 
the major dynamical classes \citep[e.g.,][]{gladman2008} have at least one large object with 
radius $r \approx$ 450--1300~km \citep[e.g.,][]{sheppard2011,petit2011}.  All-sky surveys are now 
nearly complete to an R-band magnitude $R \approx$ 21, the approximate brightness of an object with 
radius $r \approx$ 350~km and an albedo of 10\% at 42 AU \citep[e.g.,][]{schwamb2010,sheppard2011}. 
Thus, the sample of largest TNOs is probably complete at 40--50~AU.  For large TNOs with binary 
companions, period measurements yield reliable masses of roughly $1.5 \times 10^{24}$~g for Quaoar 
\citep{fraser2010a} to roughly $1.5 \times 10^{25}$~g for Pluto \citep{buie2006} and 
$1.6 \times 10^{25}$~g for Eris \citep{brown2007}.  In our calculations, this mass range corresponds 
to $r \approx$ 600--1400~km.  Thus, a typical \rmax\ = 1000$\pm$400~km matches the current population 
of the largest TNOs.

For deep surveys with sufficient observations, the luminosity function -- written as $\sigma(m)$, 
the surface number density of objects brighter than magnitude $m$ -- is reasonably well-fit by a 
double power-law,
\begin{equation}
\sigma(m) = \sigma_{m_0} 
{1 + 10^{(\alpha_L - \alpha_S)(m_{eq} - m_0)} 
\over 
10^{\alpha_L(m - m_0)} + 10^{(\alpha_L-\alpha_S)(m_{eq} - m_0) - \alpha_S (m - m_0)}}
\end{equation}
where $\sigma_{m_0}$ is a normalization factor, $m_0$ is a reference magnitude, $m_{eq}$ is the 
brightness where the two power laws merge, $\alpha_L$ is the slope of the size distribution at 
large sizes, and $\alpha_S$ is the slope of the size distribution at small sizes 
\citep[e.g.,][]{petit2006,Petit2008,fuentes2008,fuentes2009,fraser2008,fraser2009,fraser2010b,
schwamb2010,sheppard2011,petit2011}.

Converting observations of the surface number density to constraints on the size distribution 
requires additional information \citep[e.g.,][]{fraser2008,Petit2008}. If the cumulative size 
distribution and the distribution of albedos follow power laws with exponents $q$ 
(for the size distribution) and $\beta$ (for the albedo), 
\begin{equation}
q = 5 (\alpha - \beta / 2) + 1 ~ .
\end{equation}
Although TNOs display a broad range of albedos \citep[e.g.,][]{stansberry2008,brucker2009}, current data
suggest a roughly constant albedo of 2\%--20\% for $r \approx$ 30--500~km and a larger albedo of 20\% 
to nearly 80\% for $r \gtrsim$ 700~km \citep[Fig. 3 of][]{stansberry2008}. Most objects comprising the
luminosity function are faint; adopting $\beta \approx$ 0 is therefore a reasonable assumption. The
slope of the size distribution is then $q = 5 \alpha + 1$.

Recent analyses indicate a range in the slope of the luminosity function at large sizes. Several 
deep surveys yield $\alpha_L \approx$ 0.70--0.76, implying $q_L \approx$ 4.5--4.8 for $\beta$ = 0
\citep{fuentes2008,fuentes2009,fraser2009}. \citet{fraser2010b} divided their sample of the classical 
Kuiper belt into a set of cold objects with inclination $i \le$ 5\deg\ and a set of hot objects with 
$i \ge$ 5\deg. Their results suggest $\alpha_L \approx$ 0.80 ($q_L \approx$ 5) for the cold objects 
and $\alpha_L \approx$ 0.40 ($q_L \approx$ 3) for the hot objects 
\citep[see also][]{lev2001,bernstein2004,fuentes2008,fuentes2011}. 
\citet{sheppard2011} also prefer a shallower slope for the large objects. 

Other observations also distinguish the hot and cold populations within the Kuiper belt.  The largest 
cold KBOs are smaller \citep{lev2001} and have a larger fraction of binary companions \citep{noll2008} 
than the largest hot KBOs.  The cold KBOs also have larger albedos \citep{brucker2009}. Although there 
is no consensus, some results suggest that the cold KBOs are considerably redder than the hot KBOs 
\citep[e.g.,][]{tegler2000,trujillo2002,peixinho2004,gulbis2006,peixinho2008}. 

This diversity suggests that the cold and hot KBOs formed in different locations or have had different
collisional histories \citep{lev2001,malhotra1993,gomes2003,levison2003,levison2008}. In current 
dynamical models, the observed $(a, e, i)$ distributions of the hot KBOs are a result of formation 
at 15--25 AU and subsequent migration and scattering during the migration of the giant planets. In
this picture, the cold population formed at a larger distance, perhaps {\it in situ} 
\citep[see][and references therein]{batygin2011b}. 

To assign \rmax\ for the hot and cold populations, we use results from direct mass estimates of
individual KBOs \citep[e.g.,][]{grundy2009,fraser2010a,grundy2011} and deep probes of the KBO luminosity 
function \citep[e.g.,][]{fraser2009,fraser2010b,petit2011}. These analyses consistently demonstrate
that the largest hot KBOs are systematically larger than the largest cold KBOs \citep[see also][]{lev2001}. 
We infer a typical `size ratio' $\gtrsim$ 1.5--2, suggesting \rmax\ $\approx$ 1000~km for the hot objects 
and \rmax\ $\lesssim$ 600~km for the cold objects.  Although the true \rmax\ for the cold population
might be less than 600~km, results for (\qfour, \rmax) = (5, 600~km) are a reasonable proxy for any 
size in the 400--600~km ($0.4 - 1.3 \times 10^{24}$~g) range.  After deriving the constraints these 
observations place on the models, we will consider alternative models for a smaller \rmax\ =
100--200~km.

\subsection{Testing the Calculations}
\label{sec: apps-tests}

To use the observational and theoretical constraints to test our calculations, we assume that our 
\qfour\ is equivalent to the observed $q_L$.  For our purposes, the \qfour\ derived for the cold 
KBOs is identical to the \qfour\ inferred from most deep surveys. Thus, we consider the implications 
of \qfour\ = 3 and \qfour\ = 5 for \rmax\ = 600--1400~km. After a basic application of our models 
to these data in \S\ref{sec: apps-basic}, we derive a set of `success fractions,' which measure
the fraction of models that match the observations of \qfour\ and \rmax\ (\S\ref{sec: apps-succ}). 
This analysis yields two plausible models for the formation of the cold and hot TNOs: (i) ensembles
of 1~km planetesimals in a massive disk with \xm\ = 0.3--3 and (ii) ensembles of 10~km planetesimals
in a less massive disk with \xm\ = 0.03-0.3. To select the `best' model, we consider the range of
formation times at 20--50 AU (\S\ref{sec: apps-time}). These results favor ensembles of 1~km 
planetesimals as the most likely starting point for the current populations of hot and cold TNOs.

\subsubsection{Basic Tests}
\label{sec: apps-basic}

The steep size distribution of the cold Kuiper belt objects (KBOs) suggests these objects form in
a massive disk.  Although many calculations yield \qfour\ $\approx$ 5 when \rmax\ = 1000~km, the 
median \qfour\ for low mass disks is usually much smaller than 4 (Table \ref{tab: q4rmax}). Thus, 
more massive disks with $x_m \approx$ 0.3--3 are strongly favored over low mass disks with 
$x_m \lesssim$ 0.3. Because the median \qfour\ also depends on the initial planetesimal size, the 
initial disk mass required to match the observations depends on $r_0$. Larger $r_0$ favors lower mass 
disks.  For all $r_0$, models with $x_m \approx$ 0.3--3 yield good matches to the observed \qfour.

The shallow size distribution of the hot KBOs allows us to rule out a broader range of 
collision models. From Figs. \ref{fig: rmax-slope1}--\ref{fig: allrmax-slope4} and 
Table \ref{tab: q4rmax}, calculations with 100~km planetesimals never achieve \qfour\ $\lesssim$
3.5 when \rmax\ = 1000~km. Thus, these models provide a poor match to the size distribution of 
large, high inclination KBOs. Results for calculations with 10~km planetesimals are a little
more encouraging. Although calculations with a single planetesimal size fail, growth in a low
mass disk (\xm\ $\approx$ 0.03) with a range of planetesimal sizes yields \qfour\ close to 3. 

For the hot KBOs, a broad range of calculations with 1~km planetesimals result in a good match 
to the observed \qfour\ with \rmax\ = 1000~km. Models with a range of initial planetesimal sizes 
have a smaller median \qfour\ than those with a single initial size; thus, these models provide 
a better match to the observations.  Models with small initial disk masses achieve smaller 
\qfour\ than models of massive disks. Current uncertainties in the slope allow a broad range of
initial disk masses to match the observations. 

\subsubsection{Success Fractions}
\label{sec: apps-succ}

Having established rough limits on the initial disk mass and planetesimal size, we now quantify 
the range of calculations which match the observational constraints on \rmax\ and \q4. To make 
these choices, we consider ensembles of results for each set of initial conditions and fragmentation 
parameters.  From Table \ref{tab: modgrid}, there are three ensembles with $r_0$ = 10~km and 100~km, 
and five ensembles with $r_0$ = 1~km.  For each of these ensembles, we collect all pairs of 
$(r_{max}, q_4)$ and bin these data in increments of 0.1 in log \rmax. Within each \rmax\ bin,
stochastic effects produce a broad range of \qfour. Thus, we identify the fraction of calculations 
within each bin that match an adopted target \qfour\ as a function of \rmax.  To span the observed 
range, we consider two target slopes, \qfour\ = 3$\pm$0.25 and 5$\pm$0.25, for \rmax\ = 600~km, 
1000~km, and 1400~km.

To judge the uncertainties of the derived `success fractions,' we examine the variation of
the success fraction as a function of \rmax. For a single set of initial conditions 
($x_m$, $r_0$, $q_{init}$, $f_i$), the success fraction usually has an obvious trend 
(from larger to smaller success fraction or from smaller to larger success fraction) 
as a function of \rmax. 
We set the uncertainty in a single success fraction as the dispersion around this trend. 
Typically, this dispersion (and thus the adopted uncertainty) is 0.05--0.10. 

\begin{figure}
\includegraphics[width=6.5in]{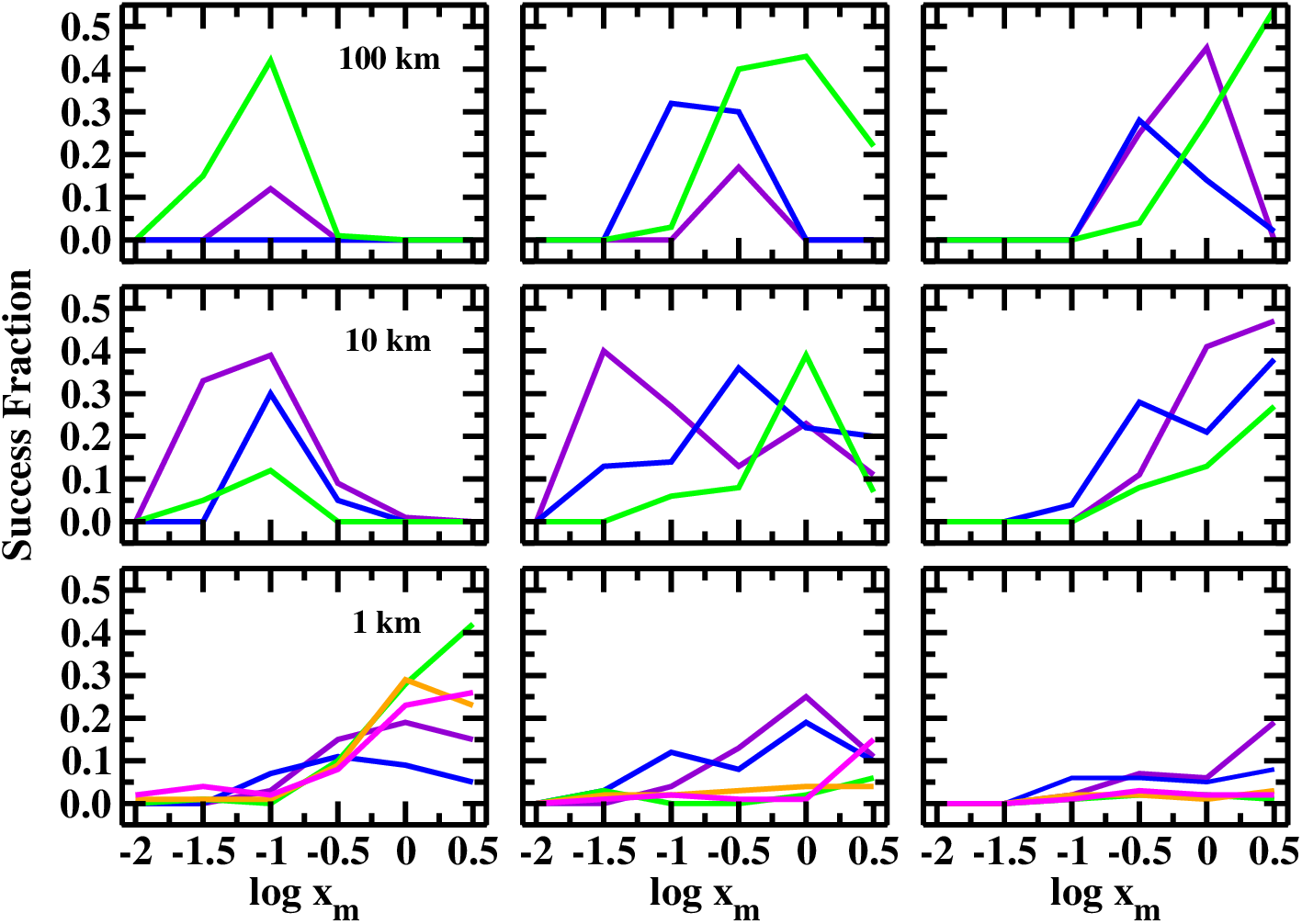}
\vskip 3ex
\caption{%
Fraction of calculations that achieve a target slope \qfour\ = 5 $\pm$ 0.25 and maximum
size log \rmax\ = 2.80 $\pm$ 0.05 (left panels), log \rmax\ = 3.00 $\pm$ 0.05 (middle panels), 
or log \rmax\ = 3.15 $\pm$ 0.05 (right panels) as a function of the initial disk mass ($x_m$). 
In each panel, the legend indicates $r_0$.  Colored lines show results for different starting 
conditions;
violet: $a$ = 15--75 AU; $n_c \propto r^{-0.5}$, $f_w$ fragmentation parameters;
blue: $a$ = 30--150 AU; $n_c \propto r^{-0.5}$, $f_w$ fragmentation parameters;
green: $a$ = 30--150 AU; $n_c \propto r^{-3}$, $f_w$ fragmentation parameters;
orange: $a$ = 30--150 AU; $n_c \propto r^{-3}$, $f_s$ fragmentation parameters;
magenta: $a$ = 30--150 AU; $n_c \propto r^{-3}$, $f_{vs}$ fragmentation parameters.
For $r_0$ = 1~km (lower panels), any range of $a$, most initial disk masses, and all 
fragmentation parameters, 10\% to 30\% of the calculations achieve the target slope 
and maximum radius; the success rate decreases slowly with \rmax.
For $r_0$ = 10~km (middle panels) and $r_0$\ = 100~km (upper panels), 20\% to 50\% of 
calculations with larger disk masses ($x_m \gtrsim$ 0.3) -- but none of the calculations with
smaller disk masses ($x_m \lesssim$ 0.3) reach the target; the success rate increases slowly 
with \rmax. 
\label{fig: prob1}
}
\end{figure}

Fig. \ref{fig: prob1} shows results for a target \qfour\ = 5$\pm$0.25. The three sets of 
vertical panels plot the fraction of successful models for \rmax\ = 600~km (left panels), 
1000~km (middle panels), and 1400~km (right panels). The three sets of horizontal panels
plot the success fraction for calculations with $r_0$ = 1~km (lower panels), 10~km (middle panels),
and 100~km (upper panels). Within each panel, colored lines show the variation of the success 
fraction as a function of the initial disk mass. Different colors indicate different initial
mass distributions and fragmentation parameters as listed in the figure caption 
(see also Table \ref{tab: modgrid}). 

Clearly, a broad range of calculations match the target slope at rates of roughly 20\% to 50\%. 
For $r_0$ = 1~km and \rmax\ = 600~km (lower left panel), calculations with any set of initial
conditions and low disk masses (\xm\ $\lesssim$ 0.1) fail to match the target slope. Models
with larger initial disk masses (\xm\ $\gtrsim$ 0.1) have larger success fractions. Within
this range of \xm, calculations with an initially large range of planetesimal sizes (green,
orange, and magenta curves) are more successful than calculations with a single planetesimal
size (violet and blue curves). As \rmax\ increases to 1000~km (lower middle panel) and to
1400~km (lower right panel), the success fraction systematically decreases. In these panels,
calculations with a single planetesimal size are somewhat more successful than those with
a range of planetesimal sizes. Once \rmax\ = 1400~km, however, few calculations successfully
match the target slope.

Calculations with larger $r_0$ also match the target slope. For $r_0$\ = 10~km and 100~km,
success fractions of 20\% to 40\% are common. Among the $r_0$ = 10~km panels, there is a clear
trend of larger success fraction with increasing initial disk mass. Although there is a small
trend of more success for calculations with a single initial planetesimal size, the trend is
much weaker than the trend with \rmax. Among the models with $r_0$ = 100~km, less (more) 
massive disks yield larger success ratios when \rmax\ is small (large). When \rmax\ is small,
calculations with a broad range of initial planetesimal sizes often match the data; results 
for a single initial planetesimal size rarely match the data. This trend reverses at large
\rmax, when calculations with a single planetesimal size are often more successful.

\begin{figure}
\includegraphics[width=6.5in]{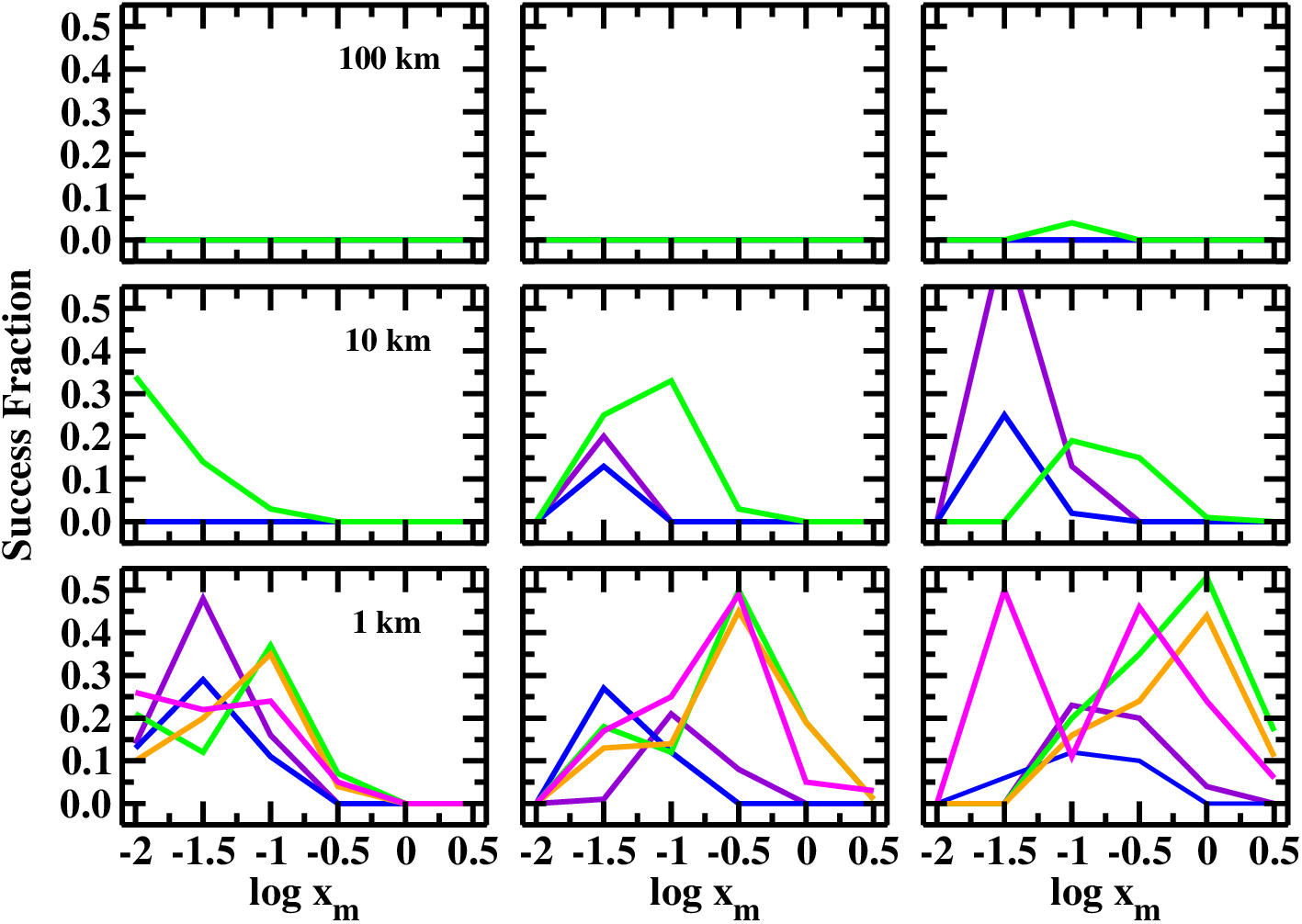}
\vskip 3ex
\caption{%
As in Fig. \ref{fig: prob1} for a target slope \qfour\ = 3 $\pm$ 0.25.
For $r_0$ = 1~km (lower panels), any range of $a$, most initial disk masses, and all 
fragmentation parameters, 20\% to 50\% of the calculations reach the target slope and 
maximum radius; the success rate increases with larger \rmax.
For $r_0$ = 10~km, calculations with large disk masses ($x_m \gtrsim$ 0.3) reach the target;
other calculations fail. 
For $r_0$ = 100~km, none of the calculations match the target.
In all calculations with $r_0$ = 10--100~km, the success rate is independent of \rmax.
\label{fig: prob2}
}
\end{figure}

Fig. \ref{fig: prob2} repeats Fig. \ref{fig: prob1} for a target \qfour\ = 3$\pm$0.25. In 
these examples, calculations with 1~km planetesimals are much more successful than those
with large planetesimals. For any \rmax\ (any vertical column of panels in the figure),
the success fractions decline monotonically from 20\%--40\% at $r_0$ = 1~km to 0\% for 
$r_0$ = 100~km. Although there is little trend in the success fraction with \rmax\ for 
calculations with $r_0$ =10--100~km, there is a modest trend for $r_0$\ = 1~km: smaller
\rmax\ requires disks with smaller masses to match the target slope. In all panels for
$r_0$ = 1~km, there is little trend with the initial size distribution of planetesimals 
or the fragmentation parameters: all of the initial conditions yield success ratios of 
20\% to 40\% for \xm\ = 0.03--3.

To refine these tests, we now include other constraints on the hot and cold populations. 
In the dynamical models, hot KBOs formed closer to the Sun and then migrated out to their 
current orbits. In the context of our coagulation models, objects closer to the Sun grow 
faster (Fig. \ref{fig: rmax1}) and produce shallower size distributions 
(Fig. \ref{fig: slope1}) over the same evolution time. Both of these features of collisional
evolution agree with the observations: the hot KBOs are larger and have a shallower size
distribution than the cold KBOs. 

To find a combination of starting conditions that match the observations, we consider models 
that match (\qfour, \rmax) = (3, 1000~km) for the hot objects and (5, 600~km) for the cold 
objects.  Using results in Figs.  \ref{fig: prob1}--\ref{fig: prob2}, we rule out any models 
starting with 100~km planetesimals. These calculations never match the shallow slope for the 
hot objects.  However, two options with 1~km and 10~km planetesimals are equally likely:
\begin{itemize}

\item 10~km planetesimals in a low mass disk (\xm\ = 0.03--0.3)

\item 1~km planetesimals in a massive disk (\xm\ = 0.3--3)

\end{itemize}
In both cases, models with a single planetesimal size or a range of planetesimal sizes
match the constraints reasonably well. Models with a single planetesimal match more often
when $r_0$ = 10~km; models with a range of sizes match more often when $r_0$\ = 1~km.

\subsubsection{The Formation Time}  
\label{sec: apps-time}

To select the `better' alternative among these two options, we add a final constraint -- the
formation time -- to our analysis. The dynamical models require TNO formation at 20--40~AU
sometime between 1--10~Myr and 500--700~Myr. The smaller of these two constraints is not
useful. Only a few models produce TNOs in 10~Myr at 20~AU; none of our calculations produce
TNOs in 10~Myr at 40--50~AU. The longer constraint is very useful. Many calculations require
more than 1~Gyr to produce 600--1000~km objects at 40--50~AU. Thus, the dynamical models rule
out these coagulation models.

\begin{figure}
\includegraphics[width=6.5in]{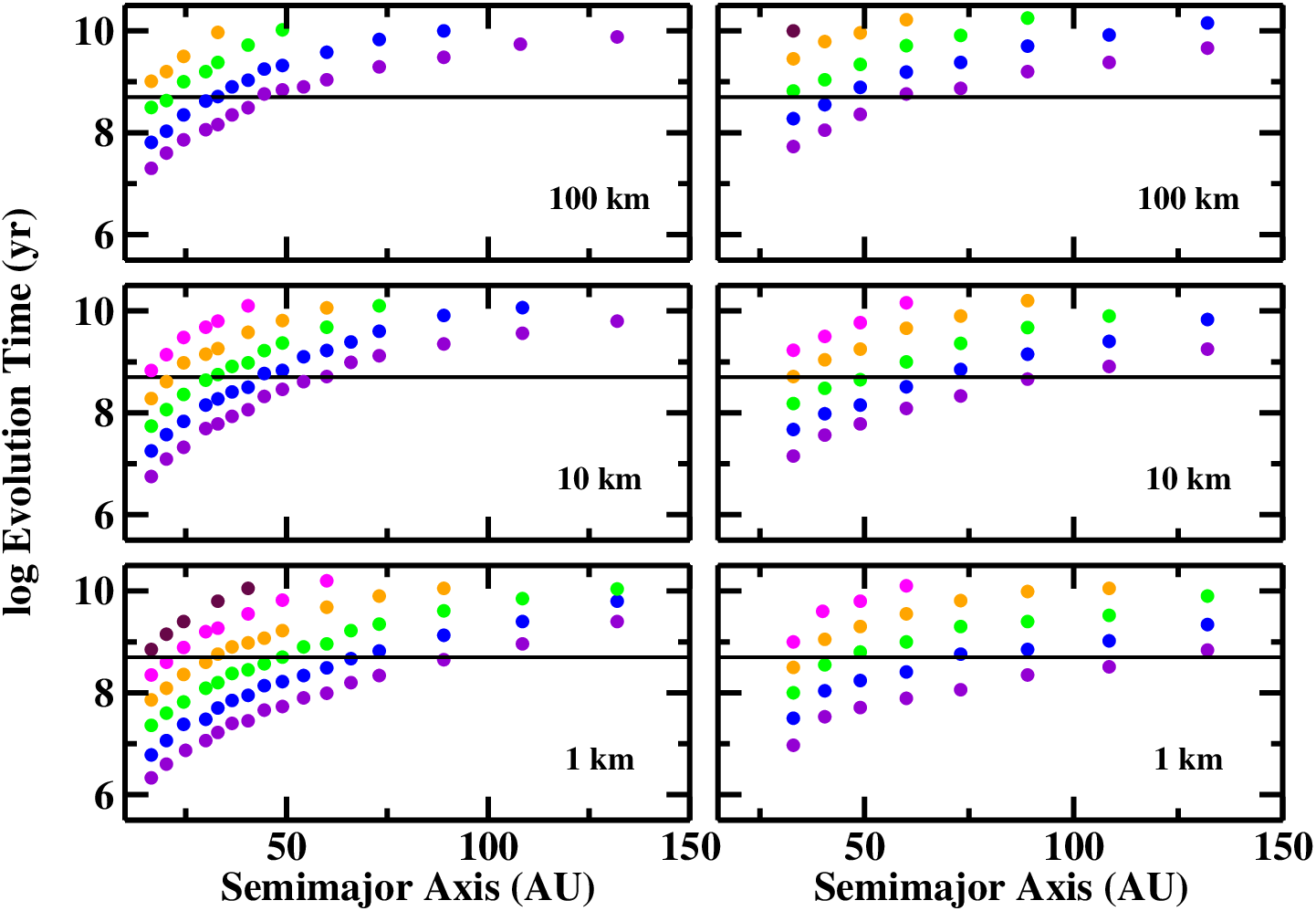}
\vskip 3ex
\caption{%
The median evolution time to reach \rmax\ = 1000~km as a function of semimajor axis
for calculations at 15--150~AU with the $f_w$ fragmentation parameters.
Left panels show results for calculations where all of the initial mass is mostly in
large planetesimals ($n_c \propto r^{-0.5}$). Right panels show results for calculations
where the initial mass is equally distributed in log mass ($n_c \propto r^{-3}$). In 
each panel, the legend indicates the initial radius of the largest planetesimal.
Colored points indicate results for initial values of $x_m$ from 
Fig. \ref{fig: rmax-slope1}. The black horizontal line illustrates the approximate time of 
the Late Heavy Bombardment \citep[e.g.,][]{tera1974,hartmann2000,stoffler2001,chapman2007}.
\label{fig: time-rmax}
}
\end{figure}

To select between the allowed combinations of $r_0$ and \xm, we consider the evolution time 
required to form 1000~km objects as a function of semimajor axis.  Fig. \ref{fig: time-rmax} 
shows results for calculations with the $f_w$ fragmentation parameters.  Results for other
calculations are similar. Individual panels plot results for $r_0$ = 1~km (lower panels), 
$r_0$ = 10~km (middle panels), and $r_0$\ = 100~km (upper panels) and two initial size 
distributions (left and right panels). Colors indicate the initial disk mass, ranging from 
$x_m$ = 3 (violet) to $x_m$ = 0.01 (maroon) as in Fig. \ref{fig: rmax-slope1}. 

These results follow standard expectations for coagulation calculations.  Formation times
always scale with the initial radius of the largest planetesimal 
\citep[e.g.,][]{liss1987,kl1998,kb2010}. As $r_0$ increases from 1~km to 100~km, the 
clusters of points shift to longer formation times.  In a disk with a surface density 
$\Sigma \propto x_m a^{-3/2}$, the formation time scales as $t_f \propto x_r^{-3} a^3$ 
\citep{gold2004,kb2008}. As $a$ increases, formation times in each panel grow rapidly.
Similarly, models with smaller \xm\ have longer formation times.

The solid horizontal line in each panel indicates the approximate end of the LHB,
500--700~Myr. Applying the additional constraint that TNOs form at $a \approx$ 
20--50 AU yields constraints on all of our calculations:
\begin{equation}
x_m \gtrsim \left\{
\begin{array}{cl}
0.01 ~ (r_0 / {\rm 1~km})^{1/2} ~ ( a / {\rm 20~AU} )^3 & ~~~{\rm all~planetesimal~sizes} , \\
0.03 ~ (r_0 / {\rm 1~km})^{1/2} ~ ( a / {\rm 20~AU} )^3 & ~~~{\rm one~planetesimal~size} .
\end{array}
\right.
\label{eq: mass-range1}
\end{equation}
When $r_0$ = 100~km, these constraints are severe: {\it in situ} formation of TNOs at 40~AU
requires $x_m \gtrsim$ 1--3. Thus, calculations with very large planetesimals are viable only 
if the initial mass of the disk exceeds the minimum mass solar nebula. For smaller $r_0$, 
calculations with smaller \xm\ yield 1000~km TNOs in 500~Myr at 40~AU.

Requiring TNOs to form in 10~Myr places more stringent limits on the initial disk mass and
the range of semimajor axes. In this case, TNOs can only form at $a \approx$ 20~AU for 
massive disks (\xm\ = 1--3) composed of 1--10~km planetesimals. Rapid formation times 
exclude all combinations of models with $r_0$ larger than 10~km.

These conclusions are independent of the fragmentation parameters. When other initial 
conditions are held fixed, formation times for icy planets with \rmax\ $\approx$ 1000~km
are similar for all three sets of fragmentation parameters. Although the size of the
largest object varies with $f_i$ (Fig. \ref{fig: allrmax-slope4}), the variation of 
\qfour\ with \rmax\ is similar for calculations with weak, strong, and very strong 
planetesimals. Thus, the formation time places strong constraints on the initial disk
mass and the initial sizes of planetesimals.

From eq. \ref{eq: mass-range1}, formation at 40--50~AU places the most severe constraints
on the initial disk mass. For 1--10~km planetesimals, we adopt the less restrictive model
with all planetesimal sizes. Then:
\begin{equation}
x_m \gtrsim \left\{
\begin{array}{cl}
0.1 - 0.2 & ~~~{\rm 1~km~planetesimals} , \\
0.3 - 0.6 & ~~~{\rm 10~km~planetesimals} .
\end{array}
\right.
\label{eq: mass-range2}
\end{equation}
Because coagulation models with 10~km planetesimals match the observed range in \qfour\ and 
\rmax\ only for $x_m \approx$ 0.03--0.3, the combined constraints of the formation time,
the size of the largest object, and the slope of the large object size distribution strongly
favor calculations with 1~km planetesimals.

\subsection{Consequences of Small Cold KBOs}

Some observations suggest the largest cold KBOs have \rmax\ $\approx$ 100--200~km instead of the 
\rmax\ = 400--600~km used in our analysis \citep[e.g.,][]{brucker2009,grundy2009,grundy2011}. 
To investigate how a smaller \rmax\ impacts our results, we now consider the range of coagulation
models which yields an observed \q\ = 5$\pm$0.25 for \rmax\ = 100--200~km. For this analysis, we 
use \qthree\ as a proxy for the slope of the size distribution for cold KBOs smaller than \rmax. 
For \rmax\ = 100--200~km, \qfour\ provides a poor measure of the slope below 100~km; \qthree\ provides
an accurate measure of the slope over 20--200~km.

To identify coagulation models which match the observed \qthree\ for a smaller \rmax, we derive 
success fractions for all sets of initial conditions in Table \ref{tab: modgrid} assuming a 
target \qthree\ = 5 $\pm$ 0.25 and \rmax\ = 100--200~km. For calculations with $r_0$ = 100~km
and \rmax\ = 100--200~km, \qthree\ is identical to the initial slope (see Fig. \ref{fig: q3rmax}). 
Because our adopted initial slope is \qthree\ = 0.17 or 3.0, our calculations never yield the 
observed \qthree\ = 5. Thus, all of our calculations with $r_0$ = 100~km fail to match the target 
slope. 

Calculations with $r_0$ = 100~km and a steeper initial size distribution probably cannot match 
the target slope.  When the initial \qthree\ $\gtrsim$ 3, most of the mass is in the smallest 
objects. For modest initial surface densities, \xm\ $\gtrsim$ 0.1, growth of these small objects 
should produce shallower slopes with \qthree\ $\approx$ 3--4.

For all other initial conditions, we derive success fractions of 10\% to 50\%. When \rmax\ is much 
smaller than 500~km, the collisional cascade has no impact on the slopes of the size distributions.
Thus, success fractions for a target (\qthree, \rmax) = (5, 100--200~km) are independent of
the fragmentation parameters. For calculations with all sizes or one size of planetesimals, 
there are weak trends of increasing success fractions in more massive disks. Lower mass disks
produce slower growth, which tends to yield shallower size distributions. However, this trend
is weak for small \rmax. 

Despite this lack of trends with $f_i$ or \xm, there are trends with the initial size and
size distribution of planetesimals. For $r_0$ = 10~km, a larger fraction of our models match 
the target \qthree\ for \rmax\ = 150--200~km (40\% to 50\%) than for \rmax\ = 100--150~km
(10\% to 30\%).  Calculations with $r_0$ = 1~km yield smaller success fractions for 
\rmax\ = 150--200~km (10\% to 35\%) than for \rmax\ = 100--150~km (35\% to 50\%). In both
cases, models with a single planetesimal size have somewhat smaller success fractions,
but the differences are not significant. 

Thus, coagulation models match the observed \qthree\ = 5 for the cold KBOs when \rmax\ = 100--200 km.
Calculations with $r_0$ = 1 km (10~km) yield the target slope more often when \rmax\ = 100--150~km
(150--200~km). With little preference for the disk mass or the fragmentation parameters, this
analysis does not identify a clear preference for $r_0$ or \xm. 

\subsection{Summary of Results}
\label{sec: apps-summary}

The most basic conclusion of our analysis is that the robust relation between \rmax\ and 
\qfour\ allows coagulation models to make clear predictions of (\qfour, \rmax) for any 
deep survey of TNOs. Unambiguous relations between \qthree\ and \rmax\ also enable clear
predictions for small TNOs. In principle, observations of (\qfour, \rmax) or (\qthree, \rmax)
can falsify current coagulation models. In practice, current observations agree with the 
predictions.

\begin{figure}
\includegraphics[width=6.5in]{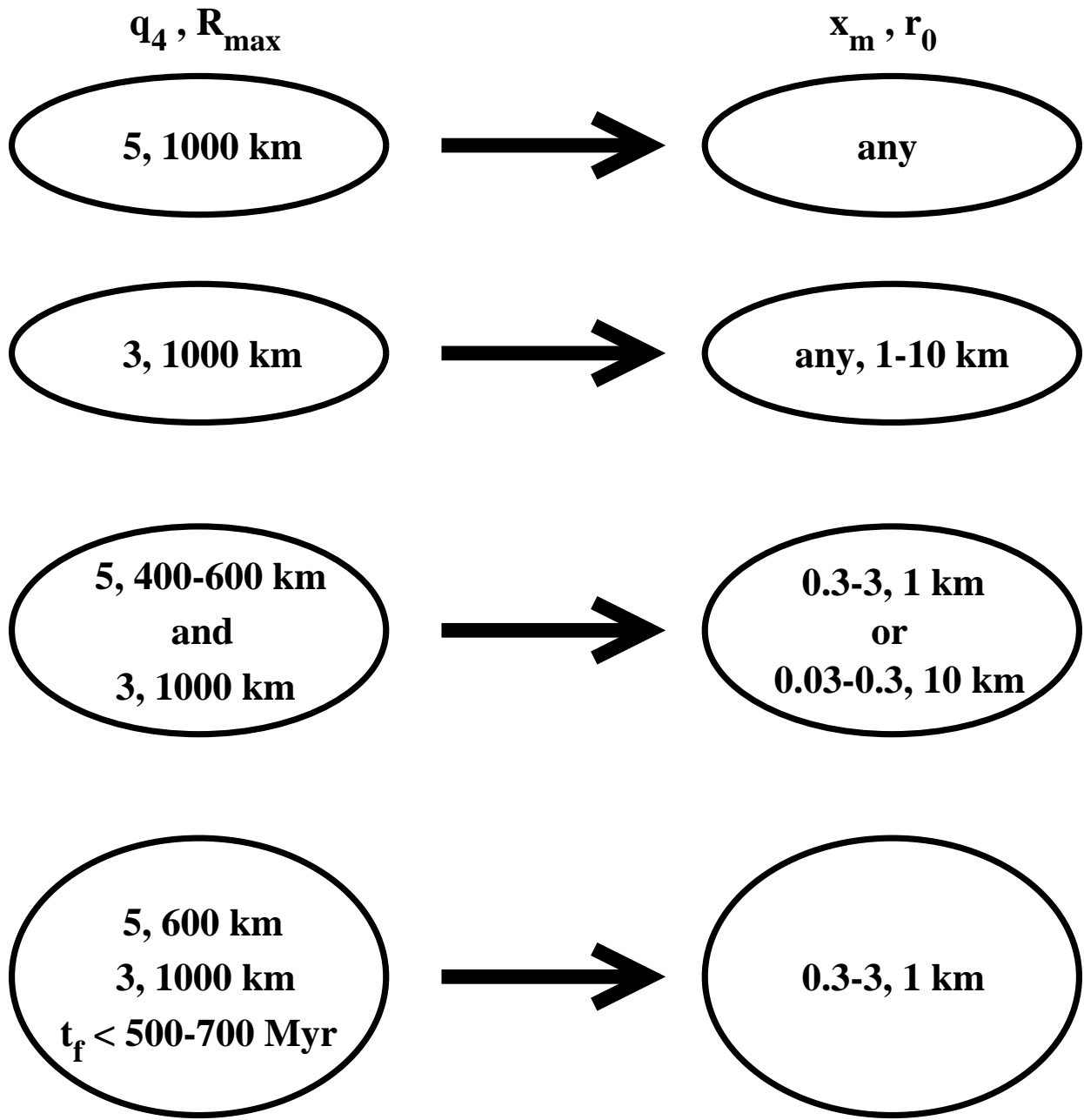}
\vskip 10ex
\caption{%
Schematic diagram for constraints on coagulation models. The left column lists target values
for \qfour\ and \rmax. The right column lists the constraints on \xm\ and $r_0$. 
Individually, measurements of \q4\ and \rmax\ for the cold (first row from top) or the hot 
(second row) population place few constraints on \xm\ or $r_0$. 
Combined (third row), these data limit coagulation models to specific combinations of
\xm\ and $r_0$.
When limits on the formation time are added (fourth row), successful models are limited
to massive disks with $r_0$ = 1~km.
\label{fig: schema}
}
\end{figure}

Using observational results from deep surveys of TNOs together with conclusions derived from 
dynamical models of the solar system, we place good constraints on coagulation models for 
the origin of TNOs throughout the solar system.  Fig. \ref{fig: schema} summarizes the main 
results of our analysis.  As we add more observational and theoretical constraints to the 
analysis, the set of coagulation calculations which `match' the constraints narrows. Thus,
a complete set of observations for all TNO classes provides a rigorous test of coagulation
models.

To summarize current limits on the models, we step through the observational and theoretical
constraints.  For a single size distribution of TNOs, we derive constraints on coagulation 
models from \qfour\ and \rmax\ (top two rows of Fig. \ref{fig: schema}).

\begin{itemize}

\item With a success rate of 20\% to 50\%, calculations with a broad range of initial 
conditions and fragmentation parameters can match a target combination (\qfour, \rmax) 
which is consistent with observations of TNOs.

\item For a specific target (\qfour, \rmax), the range of success rates allows us
to rule out some ranges of initial conditions and fragmentation parameters. 

\item When the target (\qfour, \rmax) = (5, 1400~km), calculations with large planetesimals are 
more successful than those with small planetesimals (Fig. \ref{fig: prob1}). 

\item Smaller target slopes, \qfour\ $\approx$ 3, appear to rule out calculations with 
large planetesimals (Fig. \ref{fig: prob2}). 

\item In all cases, models with large disk masses are more successful than those with 
small disk masses.  Thus, models with massive disks composed of large planetesimals are 
much more likely to match the target than low mass disks composed of small planetesimals. 

\end{itemize}

Adding in the dynamical constraint that the hot and cold populations originate in different 
locations of the protosolar disk yields more severe limits on the models. Requiring
(\qfour, \rmax) = (3, 1000~km) for the hot KBOs and (\qfour, \rmax) = (5, 400--600~km) 
for the cold KBOs (third row of Fig. \ref{fig: schema}) leads to these conclusions.

\begin{itemize}

\item Models with 100~km planetesimals cannot match the data for any set of initial conditions.

\item Models with 10~km planetesimals match the data for low mass disks, \xm\ = 0.03--0.3.

\item Models with 1~km planetesimals match the data for massve disks, \xm\ = 0.3--3.

\end{itemize}

Finally, using the constraint on formation time eliminates models with 10~km planetesimals. 
Forming TNOs in 500--700~Myr at 20--50 AU requires massive disks with \xm\ = 0.3--3 (fourth 
row of Fig. \ref{fig: schema}). Thus, we strongly favor models starting with 1~km planetesimals. 

Adopting a different observational constraint, \rmax\ = 100--200~km, for the cold KBOs results
in similar conclusions. The observed slope for the hot population, \qfour\ = 3, drives us to
models with small planetesimals in massive disks. 

\subsection{Discussion}
\label{sec: apps-disc}

Our analysis is the first attempt to apply multiannulus coagulation calculations to modern 
observations of TNOs. The results indicate that TNOs form within a massive disk composed
of small, 1~km planetesimals. This conclusion is similar to recent results on the origin
of asteroids, where an initial population of 0.1~km planetesimals seems sufficient to
explain the size distribution of asteroids for sizes of 10--100~km 
\citep[][for a different conclusion, see Morbidelli et al. 2009]{weiden2011}. Although 
we did not consider formation of TNOs from 0.1~km planetesimals, calculations with smaller 
planetesimals should yield similar results for the size distributions at 10--1000~km. Thus, 
current data are also consistent with TNO formation from 0.1~km planetesimals.

The similarity of the size distributions for the TNOs and the trojans of Jupiter 
\citep{jewitt2000,jedicke2002} and Neptune \citep{sheppard2010} suggest a common origin in
a massive disk composed of 1--10~km planetesimals \citep[for a discussion of the dynamical
origin of these populations, see][]{morby2009a}.  Unless the initial surface density of 
the disk is small (\xm\ $\lesssim$ 0.03--0.1; see eq. [\ref{eq: mass-range1}]), collisional
evolution erases the signatures of many initial conditions on 10--500~Myr timescales. Dynamical 
models for the orbital architecture require massive disks \citep{MorbiSSBN}. During the
dynamical re-arrangement of the giant planets, the large relative velocities among planetesimals 
leads to a collisional cascade which modifies the size distribution at $a \approx$ 15--30 AU
considerably for $r \lesssim$ 100~km (Figs. \ref{fig: q1rmax}--\ref{fig: q3rmax}).  

The main alternative to this picture is that the observed size distributions for the cold KBOs
and the trojans of Jupiter and Neptune are `primordial,' relatively unchanged since the initial
process that formed planetesimals from mm-sized or smaller particles \citep[e.g.,][]{sheppard2010}. 
To evaluate 
this possibility, we isolate coagulation calculations where an initial size distribution of 100~km 
and smaller objects produces 110~km or smaller objects prior to the end of LHB. For models with a 
single size or a range of sizes, calculations with \xm $\lesssim$ 0.0003~(a/40~AU)$^3$ satisfy 
this constraint. Models with larger \xm\ yield objects larger than 110~km. For primordial objects
formed {\it in situ} at $a \approx$ 20--30~AU, this limit implies an initial mass of 0.003~\mearth, 
roughly three times larger than the current mass in cold KBOs \citep[e.g.,][]{brucker2009}.
Although coagulation models do not directly preclude that cold KBOs and trojans form in a very
low mass region of the disk, it seems unlikely that dynamical mechanisms can populate three
distinct volumes of the solar system with so little reduction in the total mass \citep[e.g.,][]
{morby2005,levison2008,batygin2011a,wolff2011}. Thus, we conclude that the observed size distributions 
are not primordial.

Comparisons between our calculations and the current size distributions of TNOs make several
important assumptions. We assume that the hot and cold populations in the Kuiper belt are
representative of all TNOs. Current data do not contradict this assertion 
\citep[e.g.,][]{sheppard2010,petit2011}.  Data from ongoing and planned all-sky surveys will 
eventually yield robust estimates for (\qfour, \rmax) for all dynamical classes among the TNOs 
and will likely improve estimates for the trojans of Jupiter/Neptune.  If other dynamical classes 
of TNOs or the trojans of Jupiter/Neptune have different combinations of (\qfour, \rmax) than
analyzed here, our approach provides a framework to test coagulation models with these new data.

Detailed tests of our coagulation calculations also assume that large-scale rearrangement of
the planets in the solar system has little impact on relations between \rmax\ and \qfour.
We do not expect this evolution to produce significant changes in \qfour. Before dynamical 
interactions with the gas giants become important, gravitational focusing factors are already 
small. Collision rates are in the orderly regime, where all objects collide at comparable 
rates. Thus, \qfour\ should not evolve considerably once large-scale dynamical evolution is
important.

Throughout large-scale dynamical evolution, we expect significant changes to \qone--\qthree\ as
a result of collisional evolution. Because collisions remain in the orderly regime, these changes 
should follow the paths outlined in our discussion to Figs. \ref{fig: q1rmax}--\ref{fig: q3rmax}.
Depending on the timing of the start of the dynamical evolution, \qtwo\ and \qthree\ for dynamically
hot populations should evolve to roughly 2 or remain roughly constant at 2. If collision rates 
are large enough, \qone\ should evolve from roughly 2 to roughly 0. 

Some change in \rmax\ during the epoch of large-scale dynamical evolution seems likely.  Because 
collisions occur at high velocities, it is unlikely that \rmax\ increases. Low collision rates
imply little collisional erosion. However, the occasional high velocity, `hit-and-run' collision
\citep{asphaug2006} could remove the outer shell of material from a large object and leave the
denser core unchanged. This mechanism is one way to explain the relatively high density of
Quaoar \citep{fraser2010a}.  Dynamical evolution can also eject a few of the largest TNOs from
the solar system. Because the number of large objects is always small, a few ejections can reduce
\rmax\ and lead to variations in \rmax\ among different TNO populations.

Deriving the outcome of the dynamical evolution requires codes which combine statistical calculations 
of small objects as in this paper with the $n$-body calculations of gas giant planet formation as in 
\citet{bk2011a} and calculations of the migration and scattering of small objects as in 
\citet{charnoz2007}, \citet{levison2008,levison2010}, and \citet{bk2011b}. 
Following the relative timing of the formation of gas giants and TNOs within the same disk will 
provide better limits for \rmax, \qfour, and possible locations for the formation of TNOs within 
the protosolar disk.  Combining collisional growth with scattering will yield better predictions 
for the break radius and the slope of the size distribution below the break. Our encouraging results 
from pure coagulation models suggest a more complete model will enable a better understanding for 
the origin of TNOs and other large objects throughout the solar system.

\section{SUMMARY}
\label{sec: summary}

To develop a framework to test coagulation models for the formation of TNOs, we have analyzed 
a set of calculations in grids extending from 15--75 AU and from 30--150 AU. The calculations
consider a broad range of initial conditions for (i) the disk mass (\xm\ = 0.01--3 times the mass 
of the minimum mass solar nebula), (ii) the initial size of the largest planetesimals ($r_0$ = 1, 
10, and 100~km), (iii) the initial size distribution of planetesimals ($n_c \propto r^{-q_{init}}$,
with $q_{init}$ = 0.5 and 3), and the fragmentation parameters (weak, strong, and very strong).
Table \ref{tab: modgrid} summarizes the number of calculations with each set of parameters.

To facilitate comparisons between the calculations and observations, we focus on several observable 
features of the models. The size of the largest object in a region, \rmax, is easily derived
in the models; in any survey, the brightest TNOs are usually the largest and most massive. If the
albedos of TNOs are well-known, the slope of the size distribution follows from observations of
the TNO luminosity function. Here, we define four quantities, \qone--\qfour, which measure the
slope of the size distribution over a decade in radius. Tables \ref{tab: q4rmax}--\ref{tab: q1rmax}
list median values for these slopes as a function of \rmax\ for each set of initial conditions.

The evolution of \qone--\qfour\ as a function of \rmax\ is equivalent to the evolution as a function
of time. Thus, we treat the evolution of icy planets at all distances from the Sun on the same 
footing and derive more robust relations between the observables and the initial conditions. For 
the theoretical calculations in this paper, we infer the following conclusions 
(see also \S\ref{sec: summ-results}):

\begin{itemize}

\item For identical initial conditions, small variations in collision probabilities -- due to 
different random number seeds -- produce a broad range of outcomes for \qfour, \rmax, and 
other observables.

\item As in \citet{kb2008, kb2010}, the formation time for objects with $r$ = \rmax\ $\gtrsim$
300--1000~km depends 
inversely on the initial disk mass and with the cube of the distance from the central star, 
$t_f \propto x_r^{-3} a^3$ \citep[see also][]{saf1969,liss1987,gold2004}. 

\item Coagulation models yield a robust correlation between \rmax\ and \qfour\ as a function
of the initial conditions in the protosolar disk. This correlation allows unambiguous tests of
the models.

\item When 1000~km protoplanets grow from smaller objects, the evolution naturally leads to a
roughly power-law size distribution for objects with $r \gtrsim$ 100~km. The typical slope 
for the cumulative size distribution is \qfour\ $\approx$ 2--4 
\citep[see also][]{kl1999b,kb2004c,schlichting2011}. This slope is independent of the distance
from the Sun and the initial size distribution. Although the slope depends weakly on the 
fragmentation parameters, it is very sensitive to \xm\ and $r_0$. Larger \xm\ and larger $r_0$ yield 
steeper size distributions with larger \qfour.

\item For planetesimal sizes smaller than the initial size of the largest planetesimal ($r_0$),
the size distribution retains some memory of the initial size distribution.  However, this 
memory is limited to timescales before the onset of the collisional cascade. Once the cascade 
begins, the size distribution depends on the evolution of the largest objects and the 
fragmentation parameters for the smallest objects.

\item Once protoplanets reach radii of 500--1500~km, high velocity collisions among leftover
small planetesimals start a collisional cascade which grinds many of the leftovers to dust.
As the cascade develops, the slopes of the size distribution for 0.1--100~km objects evolve
to a standard value, $q \approx$ 2.0--2.5. Eventually, the largest protoplanets reach sizes of
2000~km or larger. High velocity collisions among the leftovers then completely shatter 1~km 
and smaller objects.  The slope of the size distribution for 0.1--1~km objects then evolves to
much smaller values, \qone\ $\approx$ 0. Slopes for larger size ranges stay roughly constant 
at \qtwo\ $\approx$ \qthree\ $\approx$ 2. 

\end{itemize}

To test the ability of coagulation models to match the observations, we focus our analysis on
the relation between \qfour\ and \rmax\ for the hot and cold populations of KBOs. The \qfour\
for the cold population and the \rmax\ for the hot population are representative of the entire 
ensemble of TNOs. Treating the two populations separately allows us to test the models as much 
as possible with current data.

Our tests are very encouraging. A broad range of coagulation calculations match the typical 
slope, \qfour\ = 5, for large objects when \rmax\ = 400--1400~km 
($4 \times 10^{23} - 1.5 \times 10^{25}$ g).  Large slopes generally favor calculations starting 
with 10--100~km planetesimals. However, smaller \rmax\ favors smaller planetesimals.
The observed slope of the hot objects, \qfour\ = 3, completely rules out models starting with 
100~km planetesimals. If TNOs must form prior to the LHB at 500--700~Myr, formation models for 
the hot objects require ensembles of 1~km planetesimals in a massive disk (\xm\ = 0.3--3).
If the largest cold objects have $r \approx$ 400--600~km ($4 \times 10^{23} - 1.5 \times 10^{24}$ g),
the LHB constraint similarly requires calculations with 1~km planetesimals. 

Combining the observations of the hot and cold populations with the constraints that they formed
in different locations of the protosolar disk prior to 500--700~Myr leaves us with one set of
starting conditions, a massive disk with 1~km planetesimals. This conclusion agrees with our 
previous result that the basic building blocks of debris disks are 1~km planetesimals \citep{kb2010}.
Although these analyses do not consider the possibility of 0.1~km or smaller planetesimals as in 
\citet{weiden2011}, the results clearly favor small planetesimals over larger planetesimals with
radii of 10--100~km.

Currently, it is difficult to make robust tests of predictions for \qone--\qthree.  Many surveys 
suggest the TNO size distribution is a double power law with a break at $R \approx$ 25--26 
\citep{fraser2008,fraser2009,fuentes2009}.  If TNOs have a typical albedo of 6\%, the break in 
the size distribution is at $r \approx$ 30~km.  
Although the uncertainties in the observations below this break are large, the slope at small sizes 
is most likely shallower than at large sizes, with $q_S \approx$ 2 a reasonable estimate.
The coagulation calculations described in this paper can match one of these two constraints. These
calculations produce a break in the size distribution at 1--3~km 
\citep[see also][]{kb2004c,gilhutton2009,fraser2009b}. 
Below this break, the collisional cascade always produces a slope of roughly 2.  Thus, our models 
in this paper match the slope below the break but not the location of the break.

\citet{kb2004c} demonstrate that stirring from the largest nearby planet sets the `break radius'
\citep[see also][]{gilhutton2009,fraser2009b}.
For TNOs, continued stirring by Neptune at 30~AU yields a break at 10--30~km, close to the lower
limit inferred from current observations. Given the uncertainties in the luminosity function at
$R$ = 25--27 and the albedos for such TNOs, coagulation models provide a satisfactory match to
the data \citep[see also][]{nesv11}. 

Improving these tests requires additional data. Among the more massive TNOs, deriving \qfour\ and
\rmax\ for all of the dynamical classes will place better limits on the ability of coagulation
and dynamical calculations to provide a unique model for the origin of the giant planets and TNOs.
As deep surveys probe larger volumes of transneptunian space, results for the observational analogs 
to \qone--\qthree\ will test the ability of coagulation models to explain size distributions for 
each class of TNOs.  For the entire ensemble of TNOs, coagulation models probably have fewer free
parameters than the set of observational constraints. Thus, TNOs should provide robust tests of
coagulation models for planet formation.

\vskip 6ex

We acknowledge generous allotments of computer time on the NASA `discover' cluster, the SI 
`hydra' cluster, and the `cosmos' cluster at the Jet Propulsion Laboratory.  We thank S.
Weidenschilling for a careful and timely review. Advice and comments from W. Fraser,
M. Geller, M. Holman, and A. Youdin also greatly improved our presentation.  
Portions of this project were supported by the {\it NASA } {\it Astrophysics Theory} 
and {\it Origins of Solar Systems} programs through grant NNX10AF35G, the {\it NASA} 
{\it TPF Foundation Science Program} through grant NNG06GH25G, the {\it NASA} 
{\it Outer Planets Program} through grant NNX11AM37G, the {\it Spitzer Guest Observer Program} 
through grant 20132, and grants from the endowment and scholarly studies programs of the 
Smithsonian Institution.

\appendix
\section{APPENDIX}

As oligarchic growth proceeds, cratering collisions gradually dominate physical interactions
among 0.1--10~km planetesimals. In this regime, collisions between equal mass objects produce
the most debris. For a power-law cumulative size distribution with $q \gtrsim$ 3, small 
planetesimals are more likely to collide with one another than with large oligarchs
\citep{kb2004a,gold2004}. Thus, cratering leads to a gradual erosion of small planetesimals
\citep[e.g.,][]{tanaka1996,koba2010}. 

During this erosive phase, collisions between equal mass objects produce a single object
with mass comparable to the target and copious debris. For these objects, the slope of the
size distribution then depends on the rate collisions convert planetesimals into debris and 
the rate of debris production over a range of sizes. To estimate the rate planetesimals are lost,
we consider the collision rate, $n \sigma v f_g$. During oligarchic growth, small planetesimals 
have similar orbital eccentricities and $f_g \approx$ 1. For collisions between equal mass 
objects, the rate small planetesimals are destroyed is $dN/dt \propto N^2 r^2$, where $N$ 
is the number of planetesimals in a size range and $r$ is a typical radius. Setting 
$N \propto r^{-q}$, 
\begin{equation}
d~{\rm log}~N / dt \approx r^{2-q} ~ .
\end{equation}
For an ensemble of planetesimals, small objects are destroyed relatively more (less) often 
when $q \gtrsim 2$ ($q \lesssim 2$). More destructive collisions among small objects reduces
the slope; fewer destructive collisions raises the slope.  Without any debris, this process 
leads to an equilibrium size distribution with $q \approx$ 2.  In our simulations, debris has 
a power-law cumulative size distribution with $q_{debris} = 2.5$. Thus, cratering produces a 
power-law size distribution with $q_{cr} \approx$ 2.0--2.5, as observed in our calculations.

As the largest objects continue to grow, collisions among small objects become more and more
catastrophic. For the fragmentation parameters we adopt, $Q_D^*$ monotonically increases with 
$r$ for $ \gtrsim$ 0.1~km. In this range, small objects fragment catastrophically before 
large objects. Thus, catastrophic fragmentation reduces the equilibrium slope from the
cratering slope of 2.0--2.5.

To provide a rough estimate of the equilibrium slope for catastrophic collisions, we consider
a collision rate where the fraction of completely destructive collisions is a decreasing function
of $r$, e.g., $f_d \propto r^{-p}$. The loss rate of planetesimals is then,
\begin{equation}
d~{\rm log}~N / dt \approx f_d ~ r^{2-q} ~ \approx r^{2-p-q}.
\end{equation}
From the expression for $Q_D^*$ (eq. [\ref{eq:Qd}]), $p \approx 1.5$.
In the absense of debris production, catastrophic collisions lead to a power-law size 
distribution with $q_{cd} \approx$ 0.5, close to the $q \approx$ 0--1 derived in our 
simulations.

\bibliography{ms.bbl}

\clearpage

\begin{deluxetable}{lccccccccccc}
\tablecolumns{12}
\tablewidth{0pc}
\tabletypesize{\footnotesize}
\tablenum{1}
\tablecaption{Grid of Calculations\tablenotemark{a} with $\Sigma \propto a^{-3/2}$}
\tablehead{
  \colhead{} &
  \multicolumn{11}{c}{Initial size of largest object ($r_0$)} \\
  \colhead{$x_m$~~~} &
  \colhead{1~km} &
  \colhead{10~km} &
  \colhead{100~km} &
  \colhead{1~km} &
  \colhead{10~km} &
  \colhead{100~km} &
  \colhead{1~km} &
  \colhead{10~km} &
  \colhead{100~km} &
  \colhead{1~km} &
  \colhead{1~km}
}
\startdata
0.01 & 7 & 7 & 7 & 7 & 14 & 7 & 7 & 8 & 10 & 14 & 15 \\
0.03 & 7 & 7 & 7 & 7 & 7 & 7 & 7 & 8 & 9 & 16 & 15 \\
0.10 & 7 & 7 & 7 & 7 & 12 & 7 & 7 & 7 & 8 & 14 & 15 \\
0.33 & 7 & 7 & 7 & 8 & 14 & 9 & 7 & 7 & 15 & 19 & 17 \\
1.00 & 7 & 7 & 7 & 8 & 16 & 7 & 14 & 7 & 13 & 15 & 16 \\
3.00 & 7 & 7 & 7 & 7 & 7 & 7 & 13 & 7 & 13 & 18 & 15 \\
\\
Notes & 
b, e, g &
b, e, g &
b, e, g &
c, e, g &
c, e, g &
c, e, g &
c, f, g &
c, f, g &
c, f, g &
c, f, h &
d, f, i \\
\enddata
\tablenotetext{a}{Number of independent calculations for each combination 
of $x_m$ and $r_0$}
\tablenotetext{b}{$a$ = 15--75 AU, $t_{max}$ = 1~Gyr, new calculations for this paper}
\tablenotetext{c}{$a$ = 30--150 AU, $t_{max}$ = 10~Gyr, calculations from \citet[][2010]{kb2008}}
\tablenotetext{d}{$a$ = 30--150 AU, $t_{max}$ = 10~Gyr, new calculations for this paper}
\tablenotetext{e}{$n_c \propto r^{-0.5}$}
\tablenotetext{f}{$n_c \propto r^{-3}$}
\tablenotetext{g}{$f_w$ fragmentation parameters \citep{lein2009}}
\tablenotetext{h}{$f_s$ fragmentation parameters \citep{benz1999}}
\tablenotetext{i}{$f_{vs}$ fragmentation parameters \citep{davis1985}}
\label{tab: modgrid}
\end{deluxetable}
\clearpage

\begin{deluxetable}{lccccccl}
\tablecolumns{8}
\tablewidth{0pc}
\tabletypesize{\footnotesize}
\tablenum{2}
\tablecaption{Median values for \qfour\ when \rmax\ = 1000~km}
\tablehead{
  \colhead{} &
  \multicolumn{6}{c}{Initial Disk Mass ($x_m$)} & \\
  \colhead{Parameter} &
  \colhead{0.01} &
  \colhead{0.03} &
  \colhead{0.10} &
  \colhead{0.33} &
  \colhead{1.00} &
  \colhead{3.00} &
  \colhead{Notes}
}
\startdata
$q_{4,3}$ & 1.78 & 2.08 & 3.62 & 4.13 & 5.06 & 5.30  & $a$ = 15--75 AU, $r_0$ = 1~km, $n_c \propto r^{-0.5}$, $f_{W}$ \\
$q_{4,3}$ & \nodata & 3.86 & 4.39 & 4.69 & 5.61 & 5.60 & $a$ = 15--75 AU, $r_0$ = 10~km, $n_c \propto r^{-0.5}$, $f_{W}$ \\
$q_{4,3}$ & \nodata & \nodata & \nodata & 5.25 & 6.27 & 6.53 & $a$ = 15--75 AU, $r_0$ = 100~km, $n_c \propto r^{-0.5}$, $f_{W}$ \\
$q_{4,3}$ & 1.83 & 2.79 & 3.64 & 5.48 & 6.31 & 6.22 & $a$ = 30--150 AU, $r_0$ = 1~km, $n_c \propto r^{-0.5}$, $f_{W}$ \\
$q_{4,3}$ & \nodata & 3.72 & 4.50 & 4.94 & 5.65 & 5.53 & $a$ = 30--150 AU, $r_0$ = 10~km, $n_c \propto r^{-0.5}$, $f_{W}$ \\
$q_{4,3}$ & \nodata & \nodata & 5.03 & 5.42 & 6.61 & 6.93 & $a$ = 30--150 AU, $r_0$ = 100~km, $n_c \propto r^{-0.5}$, $f_{W}$ \\
$q_{4,3}$ & \nodata & 2.74 & 2.94 & 2.93 & 3.53 & 4.08 & $a$ = 30--150 AU, $r_0$ = 1~km, $n_c \propto r^{-3}$, $f_{W}$  \\
$q_{4,3}$ & \nodata & 2.72 & 3.73 & 4.68 & 5.18 & 5.70 & $a$ = 30--150 AU, $r_0$ = 10~km, $n_c \propto r^{-3}$, $f_{W}$ \\
$q_{4,3}$ & \nodata & 3.52 & 4.03 & 4.74 & 5.12 & 5.63 & $a$ = 30--150 AU, $r_0$ = 100~km, $n_c \propto r^{-3}$, $f_{W}$ \\
$q_{4,3}$ & 1.85 & 2.18 & 2.58 & 3.17 & 3.55 & 4.25 & $a$ = 30--150 AU, $r_0$ = 1~km, $n_c \propto r^{-3}$, $f_{S}$ \\
$q_{4,3}$ & \nodata & 2.52 & 2.98 & 3.08 & 3.81 & 4.24 & $a$ = 30--150 AU, $r_0$ = 1~km, $n_c \propto r^{-3}$, $f_{vs}$ \\
\enddata
\label{tab: q4rmax}
\end{deluxetable}

\begin{deluxetable}{lccccccl}
\tablecolumns{8}
\tablewidth{0pc}
\tabletypesize{\footnotesize}
\tablenum{3}
\tablecaption{Median values for \qone\ when \rmax\ = 1000~km}
\tablehead{
  \colhead{} &
  \multicolumn{6}{c}{Initial Disk Mass ($x_m$)} & \\
  \colhead{Parameter} &
  \colhead{0.01} &
  \colhead{0.03} &
  \colhead{0.10} &
  \colhead{0.33} &
  \colhead{1.00} &
  \colhead{3.00} &
  \colhead{Notes}
}
\startdata
$q_{1,3}$ & 1.97 & 2.00 & 1.28 & 1.13 & 1.06 & 1.29  & $a$ = 15--75 AU, $r_0$ = 1~km, $n_c \propto r^{-0.5}$, $f_{W}$ \\
$q_{1,3}$ & \nodata & 2.23 & 2.23 & 2.24 & 2.26 & 2.26 & $a$ = 15--75 AU, $r_0$ = 10~km, $n_c \propto r^{-0.5}$, $f_{W}$ \\
$q_{1,3}$ & \nodata & \nodata & \nodata & 1.99 & 2.06 & 1.91 & $a$ = 15--75 AU, $r_0$ = 100~km, $n_c \propto r^{-0.5}$, $f_{W}$ \\
$q_{1,3}$ & 2.04 & 1.56 & 1.34 & 1.19 & 1.25 & 1.40 & $a$ = 30--150 AU, $r_0$ = 1~km, $n_c \propto r^{-0.5}$, $f_{W}$ \\
$q_{1,3}$ & \nodata & 2.25 & 2.25 & 2.25 & 2.25 & 2.26 & $a$ = 30--150 AU, $r_0$ = 10~km, $n_c \propto r^{-0.5}$, $f_{W}$ \\
$q_{1,3}$ & \nodata & \nodata & 2.06 & 1.98 & 2.16 & 1.78 & $a$ = 30--150 AU, $r_0$ = 100~km, $n_c \propto r^{-0.5}$, $f_{W}$ \\
$q_{1,3}$ & \nodata & 2.42 & 2.17 & 2.15 & 2.20 & 2.34 & $a$ = 30--150 AU, $r_0$ = 1~km, $n_c \propto r^{-3}$, $f_{W}$  \\
$q_{1,3}$ & \nodata & 2.18 & 2.28 & 2.39 & 2.18 & 2.20 & $a$ = 30--150 AU, $r_0$ = 10~km, $n_c \propto r^{-3}$, $f_{W}$ \\
$q_{1,3}$ & \nodata & 2.14 & 2.14 & 2.12 & 2.11 & 2.10 & $a$ = 30--150 AU, $r_0$ = 100~km, $n_c \propto r^{-3}$, $f_{W}$ \\
$q_{1,3}$ & 2.03 & 2.26 & 2.22 & 2.41 & 2.43 & 2.41 & $a$ = 30--150 AU, $r_0$ = 1~km, $n_c \propto r^{-3}$, $f_{S}$ \\
$q_{1,3}$ & \nodata & 2.39 & 2.51 & 2.29 & 2.31 & 2.37 & $a$ = 30--150 AU, $r_0$ = 1~km, $n_c \propto r^{-3}$, $f_{vs}$ \\
\enddata
\label{tab: q1rmax}
\end{deluxetable}

\begin{deluxetable}{lccccccl}
\tablecolumns{8}
\tablewidth{0pc}
\tabletypesize{\footnotesize}
\tablenum{4}
\tablecaption{Median values for \qtwo\ when \rmax\ = 1000~km}
\tablehead{
  \colhead{} &
  \multicolumn{6}{c}{Initial Disk Mass ($x_m$)} & \\
  \colhead{Parameter} &
  \colhead{0.01} &
  \colhead{0.03} &
  \colhead{0.10} &
  \colhead{0.33} &
  \colhead{1.00} &
  \colhead{3.00} &
  \colhead{Notes}
}
\startdata
$q_{2,3}$ & 3.25 & 3.18 & 3.63 & 3.64 & 3.61 & 3.51  & $a$ = 15--75 AU, $r_0$ = 1~km, $n_c \propto r^{-0.5}$, $f_{W}$ \\
$q_{2,3}$ & \nodata & 0.94 & 0.89 & 0.89 & 0.91 & 1.02 & $a$ = 15--75 AU, $r_0$ = 10~km, $n_c \propto r^{-0.5}$, $f_{W}$ \\
$q_{2,3}$ & \nodata & \nodata & \nodata & 2.27 & 2.24 & 2.24 & $a$ = 15--75 AU, $r_0$ = 100~km, $n_c \propto r^{-0.5}$, $f_{W}$ \\
$q_{2,3}$ & 3.30 & 3.59 & 3.61 & 3.66 & 3.50 & 3.35 & $a$ = 30--150 AU, $r_0$ = 1~km, $n_c \propto r^{-0.5}$, $f_{W}$ \\
$q_{2,3}$ & \nodata & 0.95 & 0.90 & 0.93 & 1.13 & 1.05 & $a$ = 30--150 AU, $r_0$ = 10~km, $n_c \propto r^{-0.5}$, $f_{W}$ \\
$q_{2,3}$ & \nodata & \nodata & 2.25 & 2.26 & 2.22 & 2.25 & $a$ = 30--150 AU, $r_0$ = 100~km, $n_c \propto r^{-0.5}$, $f_{W}$ \\
$q_{2,3}$ & \nodata & 4.03 & 3.70 & 3.72 & 4.00 & 3.83 & $a$ = 30--150 AU, $r_0$ = 1~km, $n_c \propto r^{-3}$, $f_{W}$  \\
$q_{2,3}$ & \nodata & 2.96 & 3.12 & 3.14 & 3.14 & 3.15 & $a$ = 30--150 AU, $r_0$ = 10~km, $n_c \propto r^{-3}$, $f_{W}$ \\
$q_{2,3}$ & \nodata & 2.56 & 2.61 & 2.69 & 2.70 & 2.73 & $a$ = 30--150 AU, $r_0$ = 100~km, $n_c \propto r^{-3}$, $f_{W}$ \\
$q_{2,3}$ & 4.44 & 4.11 & 4.00 & 3.94 & 4.00 & 3.92 & $a$ = 30--150 AU, $r_0$ = 1~km, $n_c \propto r^{-3}$, $f_{S}$ \\
$q_{2,3}$ & \nodata & 4.15 & 4.00 & 3.97 & 4.05 & 4.08 & $a$ = 30--150 AU, $r_0$ = 1~km, $n_c \propto r^{-3}$, $f_{vs}$ \\
\enddata
\label{tab: q2rmax}
\end{deluxetable}

\begin{deluxetable}{lccccccl}
\tablecolumns{8}
\tablewidth{0pc}
\tabletypesize{\footnotesize}
\tablenum{5}
\tablecaption{Median values for \qthree\ when \rmax\ = 1000~km}
\tablehead{
  \colhead{} &
  \multicolumn{6}{c}{Initial Disk Mass ($x_m$)} & \\
  \colhead{Parameter} &
  \colhead{0.01} &
  \colhead{0.03} &
  \colhead{0.10} &
  \colhead{0.33} &
  \colhead{1.00} &
  \colhead{3.00} &
  \colhead{Notes}
}
\startdata
$q_{3,3}$ & 4.17 & 4.14 & 3.85 & 3.82 & 3.71 & 3.70  & $a$ = 15--75 AU, $r_0$ = 1~km, $n_c \propto r^{-0.5}$, $f_{W}$ \\
$q_{3,3}$ & \nodata & 4.11 & 4.05 & 4.06 & 4.05 & 3.98 & $a$ = 15--75 AU, $r_0$ = 10~km, $n_c \propto r^{-0.5}$, $f_{W}$ \\
$q_{3,3}$ & \nodata & \nodata & \nodata & 0.97 & 0.93 & 0.95 & $a$ = 15--75 AU, $r_0$ = 100~km, $n_c \propto r^{-0.5}$, $f_{W}$ \\
$q_{3,3}$ & 3.87 & 3.72 & 3.67 & 3.64 & 3.70 & 3.69 & $a$ = 30--150 AU, $r_0$ = 1~km, $n_c \propto r^{-0.5}$, $f_{W}$ \\
$q_{3,3}$ & \nodata & 4.03 & 4.06 & 4.02 & 3.96 & 3.94 & $a$ = 30--150 AU, $r_0$ = 10~km, $n_c \propto r^{-0.5}$, $f_{W}$ \\
$q_{3,3}$ & \nodata & \nodata & 0.93 & 0.97 & 0.90 & 0.92 & $a$ = 30--150 AU, $r_0$ = 100~km, $n_c \propto r^{-0.5}$, $f_{W}$ \\
$q_{3,3}$ & \nodata & 3.18 & 3.43 & 3.19 & 2.93 & 3.40 & $a$ = 30--150 AU, $r_0$ = 1~km, $n_c \propto r^{-3}$, $f_{W}$  \\
$q_{3,3}$ & \nodata & 3.87 & 3.86 & 3.82 & 3.67 & 3.71 & $a$ = 30--150 AU, $r_0$ = 10~km, $n_c \propto r^{-3}$, $f_{W}$ \\
$q_{3,3}$ & \nodata & 3.16 & 3.15 & 3.16 & 3.16 & 3.17 & $a$ = 30--150 AU, $r_0$ = 100~km, $n_c \propto r^{-3}$, $f_{W}$ \\
$q_{3,3}$ & 2.91 & 3.20 & 3.34 & 3.36 & 3.25 & 3.35 & $a$ = 30--150 AU, $r_0$ = 1~km, $n_c \propto r^{-3}$, $f_{S}$ \\
$q_{3,3}$ & \nodata & 3.44 & 3.84 & 3.69 & 3.59 & 3.45 & $a$ = 30--150 AU, $r_0$ = 1~km, $n_c \propto r^{-3}$, $f_{vs}$ \\
\enddata
\label{tab: q3rmax}
\end{deluxetable}
%
\clearpage

\clearpage

\clearpage

\clearpage

\clearpage

\end{document}